\documentclass[useAMS,usenatbib]{mn2e}

\usepackage[T1]{fontenc}
\usepackage{ae,aecompl}
\usepackage{hyperref}
\usepackage{graphicx}
\usepackage{color}
\newcommand{\teff}{$T_{\rm eff}$}
\newcommand{\logg}{$\log g$}
\newcommand{\ag}{$A_{\rm G}$}
\newcommand{\mlsep}{$\langle \Delta \nu \rangle$}
\newcommand{\smsep}{$\langle \delta \nu \rangle$}
\newcommand{\numax}{$\nu_{\rm max}$}
\newcommand{\pmm}{$\pm$}
\newcommand{\msol}{M$_{\odot}$}
\newcommand{\rsol}{R$_{\odot}$}

\newcommand{\kms}{kms$^{-1}$}
\newcommand{\rad}{$R$}
\newcommand{\mass}{$M$}
\newcommand{\lum}{$L$}

\newcommand{\mhz}{$\mu$Hz}
\newcommand{\feh}{[Fe/H]}
\newcommand{\kep}{{\it Kepler}}
\newcommand{\corot}{CoRoT}

\newcommand{\gspp}{{\tt GSP\_Phot}}
\newcommand{\gsps}{{\tt GSP\_Spec}}
\title[A large seismic sample of calibration stars for Gaia]
   {A large sample of calibration stars for Gaia: \logg\ from 
   \kep\ and \corot\ fields}

\author[O. L. Creevey et al.]
   {\parbox{\textwidth}
    {O.~L.~Creevey$^{1}$\thanks{E-mail:ocreevey@oca.eu}, 
    F.~Th\'evenin$^{1}$,
    S.~Basu$^2$,
    W.~J.~Chaplin$^3$,
    L.~Bigot$^1$,
    Y.~Elsworth$^3$,
    D.~Huber$^4$,
    M.~J.~P.~F.~G.~Monteiro$^5$ and
    A.~Serenelli$^6$}\vspace{0.4cm}\\
    $^{1}$Laboratoire Lagrange, CNRS, Universit\'e de Nice
          Sophia-Antipolis, Nice, 06300, France.\\
    $^{2}$Department of Astronomy, Yale University, P.O. Box 208101,
          New Haven, CT 06520-8101, USA.\\
    $^{3}$School of Physics and Astronomy, University of Birmingham, 
          Edgbaston, Birmingham B15 2TT, UK.\\
    $^{4}$NASA Ames Research Center, Moffett Field, CA 94035, USA\\
    $^{5}$Centro de Astrof\'{i}sica and Faculdade de Ci\^encias, 
          Universidade do Porto, 4150-762 Porto, Portugal\\
    $^{6}$Institute of Space Sciences (CSIC-IEEC), Campus UAB, 
          Facultad de Ciencias, Torre C5, parell 2, Bellaterra, Spain.}

\begin{document}

\date{}

\pagerange{\pageref{firstpage}--\pageref{lastpage}} \pubyear{2002}

\maketitle

\label{firstpage}

\begin{abstract}

Asteroseismic data can be used to determine stellar surface gravities
with precisions 
of $< 0.05$ dex by using the {\it global seismic quantities} 
$\langle \Delta \nu \rangle$ and $\nu_{\rm max}$
along with standard atmospheric data such as $T_{\rm eff}$ and metallicity.
Surface gravity is also one of the {\it four} stellar properties to be derived
by automatic analyses for 1 billion stars from Gaia data 
(workpackage {\tt GSP\_Phot}).
In this paper we explore seismic data from main sequence 
F, G, K stars ({\it solar-like stars}) observed
by the {\it Kepler} spacecraft as a potential calibration source 
for the methods that Gaia will use for object characterisation ($\log g$).
We calculate $\log g$ for 
some bright nearby stars for which radii and masses are known
(e.g. from interferometry or binaries),
and using their global seismic quantities in a grid-based method, we 
determine an 
asteroseismic $\log g$ to within $0.01$ dex of the direct calculation,
thus validating {the accuracy of} our method.
We also find that errors in adopted atmospheric parameters (mainly [Fe/H]) can, 
however, cause systematic errors on the order of 0.02 dex.
We then apply our method to a list of 40 stars to deliver precise values
of 
surface gravity, i.e. uncertainties on the order of 0.02 dex, 
and we find agreement with recent literature values.
Finally, we explore the typical precision that we expect in a sample of 
400+ {\it Kepler} stars which have their global seismic quantities measured.
We find a mean uncertainty (precision) on the order of better than 0.02 dex 
in $\log g$ over the full explored range $3.8<\log g<4.6$, with the mean value
varying only with stellar magnitude ($0.01- 0.02$ dex).
We study sources of systematic errors in $\log g$ and 
find possible biases on the order of 0.04 dex, independent of 
$\log g$ and magnitude, which accounts
for errors in the $T_{\rm eff}$ and [Fe/H] measurements, as well as
from using a different grid-based method. %and different physics in grids.
We conclude that {\it Kepler} stars provide a wealth of reliable 
information that 
can help to calibrate methods that Gaia will use, in particular, for 
{source characterisation with {\tt GSP\_Phot} 
where excellent precision (small uncertainties) and 
accuracy in $\log g$ is obtained from seismic data.} 
%and this is one of the four stellar parameters to be extracted by 
%{\tt GSP\_Phot}.

\end{abstract}


\section{Introduction}

Large-scale surveys provide a necessary 
homogenous set of data for addressing key scientific questions.
Their science-driven objectives naturally determine the type of observations
that will be collected.  However, to fully exploit the survey, complementary
data, either of the same type but measured with a different instrument or 
of a different observable, needs to be obtained.
Combining data from several large-scale surveys can only result in the 
best exploitation of both types of data.

The ESA Gaia mission\footnote{http://sci.esa.int/science-e/www/area/index.cfm?fareaid=26} is due to launch in Autumn 2013.
Its primary objective is to perform a 6-D mapping of the Galaxy
(3 positional and 3 velocity data)\footnote{For stars fainter than V$\sim$17,
the radial velocities will not be available.}
by observing over 1 billion stars down to a magnitude of $V = 20$.
The mission will yield distances to these stars, and 
for about 20/100 million stars distances with precisions of less than
1\%/10\% will be obtained.

Gaia will obtain its astrometry by using broad band {\it G} 
photometry\footnote{The $G$ photometric scale is similar to $V$.}.
 %, which spans roughly XX -- XX~nm.
The spacecraft is also equipped with a spectrophotometer comprising both
a blue and a red prism BP/RP, delivering {\it colour} information.
A spectrometer will be used to determine the 
radial velocities of objects as far as $G=17$ (typical precisions range from
1--20 \kms), and for the brighter stars ($G<11$) high resolution spectra 
(R$\sim$11,500) will be available.

One of the main workpackages devoted to source characterisation is 
{\tt \gspp} whose
objectives are to obtain stellar properties for 1 billion single stars
by using 
the $G$ band photometry, the parallax $\pi$, and the spectrophotometric
information BP/RP \citep{bj10ilium}.
The stellar properties that will be derived are 
effective temperature \teff, 
surface gravity \logg, and metallicity \feh, 
and also extinction \ag\ in the astrometric $G$ band to each of the stars.
\citet{liu12} discuss several different methods that were developed to
determine these parameters using Gaia data and we refer to this paper
and references within 
for details.  
In brief, they discuss the reliability of determining 
the four parameters by using simulations, 
and in particular, they conclude that 
%Briefly, they conclude that \teff\ and \ag\ are
%strongly coupled parameters because
%the measurements have similar sensitivities to both of these, while
%\logg\ and \feh\ have weak sensitivities, implying that they are 
%naturally more difficult 
%to determine, e.g. see Figure~4 of \citealt{liu12}. In this paper, 
%the authors estimate 
they expect typical precisions in \logg\ on the order
of 0.1 - 0.2 dex for main sequence late-type stars, with 
mean absolute residuals (true value minus inferred value from simulations) 
no less than 0.1 dex for stars of all magnitudes, 
see Figure~14 and 15 of \citet{liu12}.
We note that the stellar properties derived by \gspp\  will be used 
as initial input parameters for the workpackage devoted to 
detailed spectroscopic
analysis of the brighter targets \gsps\ \citep{rb06,bij10,kordo11a,wor12}
using
the Radial Velocity Spectrometer data \citep{katz05}.
%citations here, recio-blanco etc.).
Spectroscopic determinations of \logg, \teff\, and \feh\ in general
can have large correlation errors, 
and if \logg\ is well constrained, 
\teff, \feh\ and chemical abundances can be derived 
much more precisely.
%to be subject to large correlation errors, which can 
%inhibit the determination of precise chemical abundances. % between them.
 % (see Figure ~4 of their paper).

The different algorithms for \gspp\ discussed by \cite{liu12} 
used to determine the stellar properties in
an automatic way have naturally been tested on synthetic data.  
However, to ensure the validity of the stellar properties, a set 
of 40 bright benchmark stars have been compiled and work is still 
currently underway to derive stellar properties for all of these in the most
precise, homogenous manner (e.g. Heiter, et al. in prep.).
%Among the 40 benchmark stars are well known nearby objects such as the Sun
%and $\alpha$ Cen A and B\footnote{SAM working group}.
Unfortunately, these benchmark stars will be too bright for Gaia, and so 
a list of about 500 primary reference stars has also been compiled.  
The idea is to use precise ground-based data and the most up-to-date models
(known from working with the benchmark stars) to determine
their stellar properties as accurately {(correct)} and precisely 
{(small uncertainties)} as possible. 
These primary reference stars will be observed by Gaia and thus will 
serve as a set of calibration stars.
A third list of secondary reference stars has also been compiled. 
These consist of about 5000 fainter targets.

%\subsection{Motivation II: abundance analyses}

%\subsection{Motivation III: Planet-hosting stars}

\begin{figure}
\includegraphics[width=0.5\textwidth]{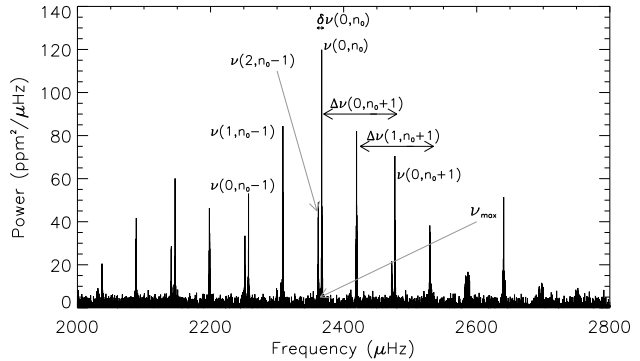}
\caption{{\label{fig:kicps} High SNR power/frequency 
spectrum of a \kep\ solar-like star KIC~6603624. 
Some individual frequencies $\nu$ are denoted with their degree $l$ and radial
order $n$ in parenthesis \citep{app12}. 
The reference radial order $n_0$ usually corresponds to orders similar to 20 for these stars.
We also show the individual large and small frequency separations, $\Delta\nu (l,n)$ and $\delta\nu(l,n)$ respectively, 
for three cases, and the approximate frequency 
corresponding to the maximum power in the spectrum, \numax.
For much lower SNR spectra, the mean value \mlsep\ of the large frequency 
separations and \numax\ can usually be determined even if individual
frequencies $\nu$ can not be resolved.}}
%The mode information is taken from \citet{app12}}}
\end{figure}

In the last decade or so, much progress in the field of asteroseismology
has been made, especially for stars exhibiting Sun-like 
oscillations.
These stars have deep outer convective envelopes where stochastic turbulence
gives rise to a broad spectrum of resonant oscillation modes 
(e.g. \citealt{ulr70}, \citealt{lei71}, 
\citealt{bg94}, \citealt{sal02}, \citealt{bc02}).  
The power spectrum of such stars can be characterised by some global seismic
quantities; \mlsep, \numax, and \smsep. The quantity \mlsep\ is the mean value
of the {\it large frequency separations} 
$\Delta\nu_{l,n} = \nu_{l,n} - \nu_{l,n-1}$ where
$\nu_{l,n}$ is a resonant oscillation frequency with degree $l$ and radial
order $n$, \numax\ is the frequency corresponding to the maximum
of the bell-shaped frequency spectrum, and \smsep\ is the mean value of 
the {\it small frequency separations} 
$\delta\nu_{l,n}=\nu_{l,n}-\nu_{l+2,n-1}$. 
Figure~\ref{fig:kicps} shows the power (frequency) spectrum of a solar-like
star, observed by the \kep\ spacecraft,  depicting these quantities.

Even when individual frequencies can not be determined from the 
frequency spectra both \mlsep\ and \numax\
can still be extracted quite robustely.
Many methods have been developed to do this using \kep-like data
\citep{bon08, sydpipeline, mosap09, hek10mnras, cam10, mat10, kar10}
and these have been compared in \citet{ver11} (see references 
therein).
The global seismic quantities have been shown
to scale with stellar parameters such as mass, radius, and 
\teff\ e.g. \citet{bg94,bed03,ste08,hub11,bed11,mig12epjwc,sil12}.
By comparing the theoretical seismic quantities with the observed ones over a 
large grid of stellar models, 
very precise determinations of \logg\ ($<0.05$ dex) 
and mean density ($<2\%$) can be obtained for main sequence F, G, K stars
\citep{bru10,met10,gai11,mat12,cre12kep}.

% (mass and radius) but are sensitive mostly
%to the ratio of these two (gravity and density):
%Although the relations are approximate, they have been found to work
%quite well, e.g. \citet{bed03,ste08,hub11,sil12}.

Of particular interest for Gaia is the {\it Kepler}\footnote{\url{http://kepler.nasa.gov}} field of view, $\sim$100
square-degrees, centered on galactic coordinates 
76.32$^{\circ}$, +13.$5^{\circ}$.
{\it Kepler} is a NASA mission dedicated to characterising planet-habitability 
\citep{bor10science}.
It obtains photometric data of approximately 150,000 stars with a typical 
cadence of 30 minutes.  However, a subset of stars (less than 1000 every 
month) acquire high-cadence data with a point every 1 minute.  This is sufficient to detect and characterise Sun-like oscillations in many stars.
\citet{ver11} and \citet{cha11science} recently showed the detections
of the global seismic quantities for a sample of 600 F, G, K 
main sequence and subgiant (V/IV) 
stars with typical magnitudes $7<V<12$, 
while both \corot\ and \kep\ have both shown their capabilities of 
detecting these
same seismic quantities in 1000s of red giants, 
e.g. \citet{hek09, kal10redgiants, 
bed10kepgiants,bau11, mos12, stello13}.

With the detection of the global seismic quantities in hundreds of 
main sequence stars (and 1000s of giants),
the {\it Kepler} field is very promising for helping to calibrate Gaia 
{\tt \gspp} methods.  In particular, the global seismic quantities deliver
one of the four properties to be extracted by automatic analysis, namely
\logg. %, and by GSP\_Spec.
\citet{gai11} studied the distribution of errors for a 
sample of simulated stars using seismic data and a grid-based method 
based on stellar evolution models.  They concluded that 
\logg\ derived from seismic properties ({\it ``seismic \logg''}) 
is almost fully independent of the input physics in the 
stellar evolution models that are used.
More recently, \citet{mm11} compared classical determinations of \logg\
to those derived alone from  a scaling relation (see Eq.[\ref{eqn:numax}]), 
and concluded 
that the mean differences between the various methods used 
is $\sim$0.05 dex, thus supporting the validity of a 
seismic determination of \logg. 
While some studies have focussed on comparing seismic 
radii or masses with alternative determinations, for example, \citet{bru10}, 
no study has been done focussing on both the {\it accuracy and precision}
%(how closely it resembles the true value)
of a seismic \logg\  using one or more grid-based
methods for stars with {\bf independently} measured radii {\it and} masses.
The accuracy and precision 
in \logg\ for these bright stars has also not been tested while considering
precisions in data such as those obtained by \kep. 
Such a study could validate 
the use of seismic data as a calibration source for Gaia.
{For the rest of this work the term {\it precision} refers exclusively
to the derived uncertainty in \logg, while {\it accuracy}
refers to how true the value is.} % and it is related to the systematic errors.

%nor has the 
%precision in \logg\ for real stars 
%been addressed using different grid-based methods, and subsequently
%tested the validity of the \kep\ field as a calibration source for Gaia.
%The precision in \logg\ for the full sample of \kep\ stars has also not 
%been yet adressed.

With these issues in mind, the objectives of this paper are to 
(i) 
test the accuracy and precision of a seismic \logg\ from a 
grid-based method using bright nearby targets for which
radii and masses have been measured (Sect.~3), %~\ref{sec:loggmethods}), 
(ii) 
determine \logg\ for an extended list of stars whose global seismic properties
and atmospheric parameters
are available in the literature using the validated method (Sect.~4), 
and %\ref{sec:extlist}), and 
(iii)
study the distribution of \logg\ and their uncertainties of over $400$ F, G, K
V/IV 
\kep\ stars
derived by a grid-based method while concluding on realistic uncertainties 
(precisions)
and possibles sources of 
systematic errors for this potential sample of Gaia calibration stars
(Sect.~5). %\ref{sec:kepstars}).
{In this work, our analysis is restricted to stars with \logg\ $>$ 3.75 
primarily due to the very limited sample of stars for which we can test
the accuracy of and validate our method.
For ease of reading, Table~\ref{tab:notation} 
summarizes the frequently used notation in this work.  }
We begin in Sect.~2 %\ref{sec:loggmethods} %In Sect.~\ref{sec:loggmethods} 
by summarising the different methods available for determining
\logg.  

\begin{table}
\begin{center}\caption{Summary of the frequently used notation in this study.\label{tab:notation}}
\begin{tabular}{llll}
\hline\hline
Notation & Definition \\%& units\\
\hline
$g$ & gravity \\
\logg & logarithm of $g$ \\
\teff & effective temperature\\
\mlsep & mean large frequency separation\\
\numax & frequency of maximum power\\
\feh & metallicity\\
%$\nu_{n,l}$ & frequency of degree $l$ and radial order $n$\\
%$\Delta\nu_{n,l}$ & individual large frequency separation \\
%&of degree $l$ and radial order $n$\\
%$\delta\nu_{n,l}$&individual small frequency separation \\
%&of degree $l$ and radial order $n$\\
$R$ & stellar radius\\
$M$ & stellar mass\\
%$L$ & stellar luminosity\\
$\sigma$ & uncertainty in \logg\\
$s$ & systematic error in \logg\\
$A_G$ & extinction in the Gaia $G$ band\\
%$T_{\rm eff pin}$ & effective temperature derived \\
%&by \citet{pin12}\\
%$T_{\rm eff irfm}$ &effective temperature derived \\
%&using the IRF method\\
r& stellar magnitude in the SDDS system\\
\hline\hline
\end{tabular}
\end{center}
\end{table}

%We also dedicate some discussion to the determination of mass, radius,luminosity, and age.

%Then in Sect.~\ref{sec:section3} 
%we apply the seismic method to a sample of bright 
%nearby stars for which accurate masses and radii (i.e. gravity) are known.
%We also simulate the determination of \logg\ for more distant stars,
%such as those measured with \corot\ and \kep, by studying the effects
%of relaxing the observational errors in our sample of stars. 
%%%We also discuss how well we determine the radius and the mass.
%In Sect.~\ref{sec:extlist} we determine \logg\ for a sample of 
%34 stars from the literature using our validated method.
%We then use the sample of \kep\ stars to study the distribution of
%uncertainties in \logg, by comparing different seismic methods and different
%input observations.  
%\red{We finally summarize the expected uncertainties and systematic errors for the 
%derivation of \logg\ as a function of magnitude and \logg.}

%\section{Accuracy of the seismic scaling relations for determining \logg}
\section{Direct methods to determine \logg\label{sec:loggmethods}}
In this section we summarise the 
various methods that are used to determine the surface gravity 
(or the logarithm of this \logg) of a star.  
Comparing each of these methods directly 
would be the ideal approach for unveiling shortcomings
in our models (systematic errors) and reducing uncertainties 
%while also increasing accuracy 
by decoupling stellar parameters.

%Model-independent methods of measuring \logg
%\red{
\subsection{Derivation of \logg\ from independent determinations of mass and radius}
The most direct method of determining \logg\ involves measuring the 
mass $M$ and radius $R$ of a star in an independent manner. 
Surface gravity $g$ is calculated using 
Newton's Law of Gravitation: $g = GM/R^2$ where $G$ is the gravitational 
constant.

\subsubsection{Mass and radius from eclipsing binary systems}
For detached eclipsing spectroscopic binaries, both $M$ and $R$ can 
be directly measured by combining photometric and radial velocity time series
\citep{rib05,cre05,cre11dscuti,hel12}.
The orbital solution is sensitive to the mass ratio and the individual
$M \sin i$ of both components, where $i$ is the inclination.  
The photometric time series displays eclipses (when the orbital plane has 
a high enough inclination) that are sensitive to $i$ and the 
relative $R$.
Once $i$ is derived, then the individual $M$ are solved.
Kepler's Law relates the orbital period of the system $\Pi$, 
the system's $M$, and 
the separation of the components.
$\Pi$ is known from either eclipse timings or observing 
a full radial velocity orbit.
Once the individual $M$ are known then $\Pi$ 
scales the system (providing the separation) and thus the individual $R$, and
$g$ can be then calculated. % using Newton's Law.

\subsubsection{Mass and radius from interferometry and global
seismic quantities}

$R$ is measured by combining the angular diameter $\theta$ 
from interferometry with the distance to the star.  
The distance (or its inverse
the parallax) has been made available using data 
from the Hipparcos satellite for stars with 
$V<\sim8$ mag \citep{per97,vanl07,ker03}.
Indirect methods also exist for determining the 
angular diameter of a star, such as combining \teff\ with measurements
of bolometric flux \citep{sil12}, 
or from calibrated relations using photometry \citep{ker04}.
%Once $R$ is known, then 
Using $R$ and \mlsep\ a model-independent mass determination can be obtained
%using $R$ and \mlsep\ and 
by using the asteroseismic relation which links mean density 
$\langle \rho \rangle$ and \mlsep
\begin{equation}
\frac{\langle \Delta \nu \rangle}{\langle \Delta \nu \rangle_{\odot}} \approx
\sqrt{\frac{\rho}{\rho_{\odot}}} = \sqrt{(M/M_{\odot})/(R/R_{\odot})^3}
\label{eqn:mlsep}
\end{equation}
where $\langle \Delta \nu \rangle_{\odot}$ % = 134.9$ \mhz\ 
is the solar value (e.g. \citealt{kb95,hub11}). % = 3,050$ \mhz\ \citep{kb95}, 
%and $R$ and $M$ are in solar units \citep{bed06,bru10,big11}.
%\red{using a seismic r, and distance from gaia, then we have theta which
%gives teff!}

\subsubsection{Mass and radius from interferometry and/or high precision
seismic data\label{sec:massradiusmodels}}

%\red{THIS TO BE REWRITTEN ...  using RADIUS and mean density}
When high signal-to-noise ratio seismic data are available,
individual oscillation frequencies (see Fig.~\ref{fig:kicps}) can be used to do detailed 
modelling, and hence determine $M$ 
(e.g. \citealt{dog10,bra11,big11,met12,mig12mnras}).
When combined with an independently measured $R$, the uncertainties in
mass $\sigma(M)$ 
can reduce to $<3\%$ \citep{cre07,baz11,hub12intast}.
If an independently determined $R$ is not available, then stellar modelling 
also yields $R$ to between 2 -- 5\% and $M$ with less precision.
%Asteroseismic data are sensitive to global stellar properties as well as 
%details of stellar interiors, e.g. sound speed profile, extent of 
%convection zones.
However, this method depends on the physics in the interior stellar models
unlike the methods mentioned above, and 
using different input physics may result in different values of mass.
Typical uncertainties/accuracies in $M$ does not usually
exceed about 5\% for bright targets when more constraints
are available, and this translates to less
than a 0.02 dex error in \logg\ for stars that we consider in this work.
As $g$ is sensitive to seismic data, then $R$ and $M$ are correlated and \logg\
can de determined with a precision of $\sim0.02$ dex with a very slight 
dependence on the physics in the models, e.g. \citet{met10}.

\subsection{Spectroscopic determinations of \logg}
The surface gravity of a star is usually derived from an atmospheric 
analysis with spectroscopic data (e.g. \citealt{the92,bru10,leb12,sou12}). 
There are two usual approaches for deriving atmospheric parameters 
(\logg, \teff, and \feh).
The first approach is based on directly comparing a library of 
synthetic spectra with the observed one, usually in the form of 
a best-fitting approach.  
A shortcoming of this method 
is that combinations of parameters can produce similar synthetic spectra so
that many correlations between the derived parameters exist. 
The more classical method for determining atmospheric parameters relies on
measuring the equivalent widths of iron lines (or other chemical species).
This method assumes local thermodynamic equilibrium (LTE) 
and requires model atmospheres.
Once \teff\ is determined (by requiring that the final line abundance is 
independent of the excitation potential or for stars with \teff$>5000$ K,
measuring the Balmer H line profiles), 
then  % and the equivalent widths of the Fe-I lines.  
$g$ is the only parameter controlling the ionisation balance of 
a chemical element in the photospheric layers, which acts on
the recombination frequency of electrons and ions.  
$g$ is then determined by requiring a balance between different
ionized lines e.g. Fe-I and Fe-II.
Spectroscopically determined \logg\ can have large systematic errors especially
for more metal-poor stars where NLTE effects must be taken into account.  
This can change \logg\ by $\sim$0.5 dex, e.g. \citet{the92}.

%of the plasma which are known as the Boltzman and Saha equations.
%The main assumption of the success of that process
%to derive \logg\ is the LTE. If this assumption is not fully satisfied,
%this could lead to an error in the value of the surface gravity. 
%This is known as non-LTE effects that is suspected to be real and strong in
%metal-poor stars
%(Th\'evenin & Idiart 1999, Fabrizzio et al. 2012). Then the derived log
%g (LTE) is lower when compared to Hipparcos log g (see figure in Nissen
%and Schuster 1997, and correct when introducing non-LTE corrections (see
%figure 8 in Thévenin & Idiart 1999).

%ionisation balance, $g$-sensitive lines

\subsection{\logg\ derived from Hipparcos data}

An alternative method for determining \logg\ relies on knowing the distance $d$
(in pc) to the star from astrometry, its bolometric flux 
(combining these gives the luminosity of a star) and \teff.  
Then substituting $M$ and \logg\ for $R$ in Stefan's Law, 
one obtains the following 
$\log g=-10.5072+\log M + 0.4(V_0+{\rm BC}_V)-2 \log d + 4 \log T_{\rm eff}$, 
%\red{$\log g=\log M + 0.4(V_0+{\rm BC}_V)-2 \log d + 4 \log T_{\rm eff} - 10.7089$, }
where $V_0$ is the de-reddened $V$ magnitude and BC$_V$ the bolometric
correction in $V$, and $M$ (in \msol) a fixed value 
that is estimated from evolution models, e.g. 
\citealt{the01,bar03}.
This {\it Hipparcos} \logg\ is often used as a fixed parameter for
abundance analyses of stars.  Typical uncertainties are no less
than 0.08 dex where especially the error in $M$ is large, and errors
from $d$ and \teff\ are not insignificant.
We note that Gaia will deliver unmistakeably accurate distances for 
much fainter stars and these will provide a much improved \logg\ using 
this method for individual stars. 
%, although interstellar extinction will decrease the accuracy.

\subsection{\logg\ from evolutionary tracks}

\subsubsection{Classical constraints in the H-R diagram}

When H-R diagram constraints are available (\teff, $L$, metallicity) then 
stellar evolution tracks can be used to provide estimates of some 
of the other parameters of the star, e.g. mass, radius, and age.
While correlations exist between many parameters, e.g. 
mass and age, these correlations also allow us to derive certain information
with better precision, e.g. mass and radius gives $g$.  
Exploring a range of models that pass through the error box thus allows us to
limit the possible range in \logg\ (e.g. \citealt{cre12chara}).

\subsubsection{Grid-based asteroseismic \logg}

%Asteroseismic data are sensitive to the global properties of stars 
%(e.g. radius, mass, and \teff) 
%as well as details of the stellar interiors e.g. sound speed profile, extent 
%of convection zones.  The typically observed range of oscillation frequencies
%gives information mainly regarding size and evolution stage of the star, or 
%density and gravity, while the distances between individual frequencies is
%most sensitive to the different layers in the stellar interior.  

Apart from performing detailed modelling using asteroseismic data
(Sect.~\ref{sec:massradiusmodels}), 
one can rely on grids of stellar models to estimate stellar properties such
as mass and radius with precisions on the order of 2--12\%.
However, because asteroseismic data are extremely sensitive to the 
ratio of these two parameters, then very precise determinations of \
\logg\ and $\langle \rho \rangle$ can be obtained in an almost 
model-independent manner (e.g. \citealt{gai11}, in Sect.~\ref{sec:yrec} 
we address this issue).
Such a {\it grid-based asteroseismic \logg} can be obtained in the following
manner:
a large grid of stellar models that spans a wide range of 
mass, age, and metallicity is constructed.
Each model in the grid has a corresponding
set of theoretical observables, such as \logg, \teff, 
individual frequencies $\nu_{n,l}$, \mlsep\ and \numax.
A scaling relation is used to obtain \numax:
%is the standard scaling relation used:
%he quantity \numax\ is calculated from  
\begin{equation}
\frac{\nu_{\rm max}}{\nu_{\rm max, \odot}} \approx \frac{(M/M_{\odot})}{(R/R_{\odot})^2\sqrt{T_{\rm eff}/T_{\rm eff,\odot} }} 
\label{eqn:numax}
\end{equation}
where $\nu_{\rm max, \odot}$ is the solar value (e.g. \citealt{van68,tas80,kb95}).
The quantity \mlsep\ can be obtained either from Eq.~\ref{eqn:mlsep} 
%can be used to calculate \mlsep\ or it can 
or calculated directly from the oscillation frequencies derived from
the structure model. 
%and in Section~3 we detail the method used in this paper.
Differences, however, on the order of 2\% in \mlsep\ 
may be found by adopting one or the other
method (e.g., \citealt{whi11}, see also \citealt{mos12b}).
%Although, recently \citet{mos12b} showed that discrepancies on the order of 2\% in \mlsep\ may be found by adopting
%one or the other method.
A set of input observed data, e.g. \{\mlsep, \teff, and \feh\}, is compared
to the theoretical one, and the models that
give the best match to the data are selected and the value of \logg\ and its
uncertainty is derived using these best selected models.
Such a precise value for this seismic \logg\ comes, in fact, from the very close relation between \numax\ and the cut-off frequency,
and recent work has made progress on understanding this relation \citep{bel11}.
%this has only recently been justified theoretically \citep{bel11}.  
Several seismic grid-based approaches for determining stellar properties
 have been discussed and applied 
in the recent literature \citep{ste09,qui10,basu10,gai11,cre12kep,met12,sil12}, 
%These methods may use either the individual frequencies or Eq.~\ref{eqn:mlsep} to calculate
%\mlsep, 
%give a refined version of this relation and they do in
%fact show that both ways of calculating \mlsep\ can result in 
%a difference of about 2\%.
and for the rest of this paper we use the term {\it seismic \logg} to refer
specifically to the grid-based method for determining \logg.

%This method has been used for some \kep\ stars and typical precisions in
%\logg\ are less than $\sim0.04$ dex \citep{met10,cre12kep}.

\section{Comparison of direct and seismically determined \logg.
\label{sec:section3}} 

\subsection{Observations and direct determination of \logg}

In order to test the reliability of an asteroseismically determined \logg,
the most correct method is to compare it to \logg\ derived from
mass and radius measurements of stars in eclipsing binaries 
(Sect.~2.1.1), apart from the Sun.  
Unfortunately the number of stars whose masses and radii are known from
binaries, where seismic data are also available, is quite limited.  
For this sample of stars we have $\alpha$ Cen A and B, and Procyon.
Following this, %the next possible model-independent determinations of 
%\logg\ 
we rely on the combination of asteroseismology and interferometry to 
determine \logg, and
use the scaling relation which links density to the observed properties
of \mlsep\ and $R$ to provide an independent $M$ measurement (Sect.~2.1.2).  
However, since this scaling relation is used explicitly 
in the grid-based method,
we have opted to omit stars where this method provides the mass, except 
%this method to provide the necessary data, except
for the solar twin 18 Sco, which was included because of its similarity to
the Sun.
To complete the list of well-characterised stars we have then chosen some 
targets for which an interferometric $R$ has been measured and detailed
seismic modelling has been conducted to determine the star's $M$ by several
authors (Sect.~2.1.3).
The stars which fall into this group are HD\,49933 and $\beta$ Hydri.
The seven stars are listed in order of \logg\ in Table~\ref{tab:refstars} along
with  \mlsep, \numax, \teff, \feh, $M$, and $R$. 
When several literature values are available these are also listed.
The final column in the table gives the {\it direct} value of \logg\ as derived
from $M$ and $R$.   
For HD\,49933 and $\beta$ Hydri 
we adopt the weighted mean values of \logg\ which are 4.214 and 3.958 dex
respectively (see Table~\ref{tab:refstars}) {and 
these are summarized in column 2 of Table~\ref{tab:radexgmr}.}
For the rest of the paper we refer to these determinations of \logg\ as 
the {\it direct} determinations.
%For clarity, in Table~\ref{tab:reflogg} we summarize the values that we adopt
%in this paper for $R$, $M$, and \logg\ for the seven stars.

\begin{table*}
\begin{center}
\caption{Properties of the reference stars\label{tab:refstars}}
\begin{tabular}{lccccccccccccccccccc}
\hline\hline
Star & \mlsep & \numax & \teff\ & \feh & \rad & \mass & \logg\\
&($\mu$Hz) & (mHz) & (K) & (dex) & (\rsol)&(\msol)&(dex)\\
\hline
$\alpha$CenB &161.5 \pmm\ 0.11$^{1a}$&4.0$^{1a}$&5316 \pmm\ 28$^{1b}$&+0.25 \pmm\ 0.04$^{1b}$&0.863 \pmm\ 0.005$^{1c}$ &0.934 \pmm\ 0.006$^{1d}$&4.538 \pmm\ 0.008\\
&&&&&0.863 \pmm\ 0.003$^{1e}$\\
18 Sco & 134.4 \pmm\ 0.3$^{2a}$ & 3.1$^{2a}$ & 5813 \pmm\ 21$^{2a}$ & 0.04 \pmm\ 0.01$^{2a}$ & 1.010 \pmm\ 0.009$^{2a}$ & 1.02 \pmm\ 0.03$^{2a}$ & 4.438 \pmm\ 0.005\\
Sun & 134.9 \pmm\ 0.1$^{3a}$ & 3.05$^{3b}$ & 5778 \pmm\ 20$^{3c}$ & 0.00 \pmm\ 0.01 & & & 4.437 \pmm\ 0.002$^{3d}$\\
$\alpha$CenA&105.6$^{4a}$&2.3$^{4a}$&5847 \pmm\ 27$^{1b}$&+0.24 \pmm\ 0.03$^{1b}$&1.224 \pmm\ 0.003$^{1c}$&1.105 \pmm\ 0.007$^{1d}$&4.307 \pmm\ 0.005 \\
 &&2.4$^{4b}$&&&\\
%HD\,49933 &85.66 \pmm\ 0.18$^{5a}$&1.8$^{5a}$&6500 \pmm\ 75$^{5b}$ &-0.35 \pmm\ 0.10$^{5b}$ &1.46 \pmm\ 0.05$^{5b}$ & &4.24$^{5b}$\\
%&&1.657 \pmm\ 0.028$^{5c}$ &&&1.49$^{5b}$ & 1.325$^{5b}$ & 4.22$^{5b}$\\
HD\,49933 &85.66 \pmm\ 0.18$^{5a}$&1.8$^{5a}$&6500 \pmm\ 75$^{5b}$ &--0.35 \pmm\ 0.10$^{5b}$ &1.49$^{5b}$ & 1.325$^{5b}$  &4.23 \pmm\ 0.02$^{5b}$\\
&&1.657 \pmm\ 0.028$^{5c}$ \\    %&&&& 4.22$^{5b}$\\
&&&6640 \pmm\ 100$^{5d}$&--0.38$^{5d}$&1.42 \pmm\ 0.04$^{5d}$ & 1.20 \pmm\ 0.08$^{5d}$ & 4.212 \pmm\ 0.039\\
&&&&&1.39 \pmm\ 0.04$^{5e}$ & 1.12 \pmm\ 0.03$^{5e}$ & 4.201 \pmm\ 0.027\\ 
&&&&&1.44 \pmm\ 0.04$^{5e}$ & 1.20 \pmm\ 0.03$^{5e}$ & 4.200 \pmm\ 0.027\\
Procyon & 55.5 \pmm\ 0.5$^{6a}$ & 1.0$^{6b}$ & 6530 \pmm\ 90$^{6c}$ & --0.05 \pmm\ 0.03$^{6d}$ & 2.067 \pmm\ 0.028$^{6e}$ & 1.497 \pmm\ 0.037$^{6f}$ & 3.982 \pmm\ 0.016\\
$\beta$Hydri&57.24 \pmm\ 0.16$^{7a}$ &1.0$^{7a}$&5872 \pmm\ 44$^{7b}$ &--0.10 \pmm\ 0.07$^{7c}$ & 1.814 \pmm\ 0.017$^{7b}$ & 1.07 \pmm\ 0.03$^{7b}$ & 3.950 \pmm\ 0.015\\
&&&&&&1.04$^{7d}$&3.938 \pmm\ 0.015\\
&&&&&&1.082$^{7e}$&3.955 \pmm\ 0.015\\
&&&&&&1.10$^{+0.04,7f}_{-0.07}$ & 3.962 \pmm\ 0.030\\
&&&5964 \pmm\ 70$^{7g}$&--0.03 \pmm\ +0.07$^{7g}$ & 1.69 \pmm\ 0.05$^{7g}$ & 1.17 \pmm\ 0.05$^{7g}$&4.02 \pmm\ 0.04$^{7g}$\\
\hline\hline
\end{tabular}
\end{center}
References:
$^{1a}$\citet{kje05}, $^{1b}$\citet{pdm08}, $^{1c}$\citet{ker03}, $^{1d}$\citet{pou02}, $^{1e}$\citet{bigot06alphacenb},
$^{2a}$\citet{baz11}, 
$^{3a}$Taking the average of Table~3 from \citet{tf92}, $^{3b}$\citet{kb95}, $^{3c}$\citet{gs98}, $^{3d}$1\msol=1.98919e30 kg, 1\rsol=6.9599e8 km, $G=6.67384e11$ m$^3$ kg$^{-1}$ s$^{-2}$(NIST database),
$^{4a}$\citet{bc02}, $^{4b}$\citet{qui10}, 
%$^{5a}$Using the $l=0$ modes with Height/Noise$>$1 from Table 1 of \citet{ben09}, $^{5b}$\citet{kal10} $Z=0.008 \pm 0.002$ is referenced, $^{5c}$\citet{gru09}, $^{5d}$\citet{big11}, $^{5e}$\citet{cre11},
$^{5a}$Using the $l=0$ modes with Height/Noise$>$1 from Table 1 of \citet{ben09}, $^{5b}$\citet{kal10} \logg\ is the mean and standard deviation of the values cited from their Table 1, $^{5c}$\citet{gru09}, $^{5d}$\citet{big11}, $^{5e}$\citet{cre11},
$^{6a}$\citet{egg04}, $^{6b}$\citet{mar04}, $^{6c}$\citet{fuh97}, $^{6d}$\citet{all02}, $^{6e}$\citet{ker04procyon}, $^{6f}$\citet{gir00},
$^{7a}$\citet{bed07}, $^{7b}$\citet{nor07}, $^{7c}$\citet{bru10}, $^{7d}$\citet{bra11}, $^{7e}$\citet{dog10}, $^{7f}$\citet{fm03}, $^{7g}$\citet{das06}
\end{table*}

%\begin{table}
%\begin{center}
%\caption{Observed $R$, $M$, and \logg\ for the reference stars\label{tab:reflogg}}
%\begin{tabular}{lccccccccccccccccccc}
%\hline\hline
%Star & $R$ & $M$ & \logg\\
%& (dex) &(\rsol)&(\msol)\\
%\hline
%$\alpha$ Cen B & 0.863 \pmm\ 0.005 & 0.934 \pmm\ 0.006& 4.538 \pmm\ 0.008 \\
%18 Sco& 1.010 \pmm\ 0.009 & 1.02\pmm\ 0.03 & 4.438 \pmm\ 0.005 \\
%Sun  & 1.000 \pmm\ 0.010 & 1.000 \pmm\ 0.010& 4.437 \pmm\ 0.002\\
%$\alpha$ Cen A  & 1.224 \pmm\ 0.003 & 1.105 \pmm\ 0.007& 4.307 \pmm\ 0.005\\
%HD~49933  & 1.42 \pmm\ 0.04$^a$ & 1.20 \pmm\ 0.08$^a$& 4.218 \pmm\ 0.011$^b$\\
%Procyon  & 2.067 \pmm\ 0.028 & 1.497 \pmm\ 0.037& 3.982 \pmm\ 0.016\\
%$\beta$ Hydri  & 1.814 \pmm\ 0.017 & 1.086 \pmm\ 0.019$^c$& 3.952 \pmm\ 0.008$^a$\\
%\hline
%\end{tabular}
%\end{center}
%$^a$The interferometric radius and corresponding asteroseismic 
%mass is used as reference.
%$^b$The weighted mean value is calculated from Table~\ref{tab:refstars} while adopting a 0.02 dex error for the measurements
%without a cited error.
%$^c$The mean and standard deviation of the mass from Table~\ref{tab:refstars} is
%used.
%\end{table}

\subsection{Seismic method to determine \logg\label{sec:radex10}}
We use the grid-based method, RadEx10, to determine an asteroseismic value 
of \logg\ 
\citep{cre12kep}. 
The grid was constructed using the  ASTEC stellar evolution code 
\citep{jcd08astec} with the following configuration:  
the EFF equation of state \citep{egg73} without Coulomb corrections,
the OPAL opacities \citep{ir96} supplemented by Kurucz opacities
at low temperatures, solar mixture from \citet{gn93}, 
and nuclear reaction rates from \citet{bp92}. 
Convection in the outer convective envelope is 
described by the mixing-length theory of \citet{bv58} and this is 
characterised
by a variable parameter $\alpha_{\rm MLT}$ (where $l=\alpha_{\rm MLT}H_p$, 
$l$ is the mixing-length and $H_p$ the pressure scale height).
When a convective core exists, there is an overshoot layer which 
is also characterised by a convecctive core overshoot parameter
$\alpha_{\rm ov}$ and this is set to 0.25
(an average of values used recently in the literature).  We also 
ignore diffusion effects, although we note that for accurate masses and 
ages this needs to be taken into account. %we need to include this ingredient. 

%  For a correct determination of mass and age, diffusion
%is an important ingredient in stellar models.

The grid considers models with masses $M$ from 0.75~--~2.00~\msol\ 
%\ in steps of 0.05 \msol.
%This quantisation in mass is not significant as stellar properties are generally
%linear for main sequence stars in this mass range, and in \citep{cre12proc}
%we show that the masses for these stars are well determined.
and ages $t$ from ZAMS to subgiant.
The initial metallicity 
$Z_{\rm i}$ spans 0.003 -- 0.030 in steps of $\sim0.003$, while 
initial hydrogen abundance 
$X_{\rm i}$ is set to 0.71: this corresponds to an initial He
abundance  $Y_{\rm i} = 0.260 - 0.287$. %%0.263-0.283
 The mixing length parameter $\alpha_{\rm MLT} = 2.0$ is used, which
 was obtained by calibration with solar data. 
We note that to derive other stellar properties, it is very important to 
allow $\alpha_{\rm MLT}$ to vary e.g. \citet{cre12chara, bon12}.
%, and varying 
%$\alpha_{\rm ov}$ will change the determination of the age of the star.

{To obtain the grid-based model stellar properties, e.g. \logg, $M$, and $t$,
%(\logg, \mass, \rad, luminosity \lum, \age)
we perturb the set of input observations by scaling its input error with
a number drawn randomly from a normal distribution and add this to the 
input (observed) value.
We compare the perturbed observations to the model ones and
select the model that matches best.
This is repeated 1,000 times to yield model distributions of 
best-matching stellar parameters.
In this method, \mlsep\ is calculated using Eq.~\ref{eqn:mlsep} and 
the input observations consist primarily of \mlsep, \numax, \teff, and \feh, 
although other inputs are possible, for example, \lum\ or \rad.
The model distributions are then fitted to a Gaussian distribution and the 
fitted stellar property and its uncertainty ($\sigma$) are defined
as the central value and the standard deviation $\sigma$ of the Gaussian fit.
%The stellar parameters and  uncertainties are defined as the mean value of the 
%fitted parameter from 10,000 realizations, with the standard deviations % of the
%defining the 1$\sigma$ uncertainties.
In this work we consider just the derived value of \logg.}

\subsection{Analysis approach\label{sec:analysisapproach}}
%Determination of \logg: how well does seismic \logg\ fare?}

%\subsection{Seismic analysis}

We determine a seismic \logg\ for the reference stars using the method
explained above.  
We consider different sets of 
input data in order to test the effect of the different observations on 
the accuracy (and precision) of a seismic \logg:\\
(S1) \{\mlsep, \numax, \teff, \feh\},\\
(S2) \{\mlsep, \numax, \teff \}, and \\
(S3) \{\mlsep, \teff \}, \\
%(S4) \{\numax, \teff \}, and \\
%(S5) \{\mlsep, \numax \}.
%%%(S4) \{\mlsep, \teff, \feh \}.\red{should do no fe/h, and s5=numax,teff}\\

For the potential sample of Gaia calibration stars, \feh\ will not always
be available, and in some cases, \numax\ is difficult to determine for very low S/N detections.
%possible to determine because the oscillation spectrum may display a double-Gaussian form instead of a single Gaussian. 
%Additionally, problems with \teff\ determinations is not an unlikely source
%of error.  For these reasons we include S2, S3, and S5.

The observational errors in our sample are very small due to the 
brightness (proximity) of the star, so we also derive an 
asteroseismic \logg\ while considering observational errors that we expect
for \kep\ stars (see \citealt{ver11} and Figure \ref{fig:cd}).
We consider three types of observational errors while repeating
the exercise:\\
(E1) the true measurement errors from the literature,\\
(E2) typically ``good'' errors expected for these stars, 
i.e. $\sigma($\mlsep) = 1.1 \mhz, $\sigma(\nu_{\rm max}) = 5$\%, 
$\sigma(T_{\rm eff}) = 70$ K, and $\sigma({\rm [Fe/H]}) = 0.08$ dex
(see Sect.~\ref{sec:kepobs}), and \\
(E3) ``not-so-good'' measurement errors, primarily considering the fainter
targets (V$\sim$11,12), 
i.e. $\sigma($\mlsep) = 2.0 \mhz, $\sigma(\nu_{\rm max}) = 8$\%, 
$\sigma(T_{\rm eff}) = 110$ K, and $\sigma([{\rm Fe/H}]) = 0.12$ dex.

\subsection{Seismic versus direct \logg\label{sec:seismicdirect}}
\begin{figure*}
\includegraphics[width=0.98\textwidth]{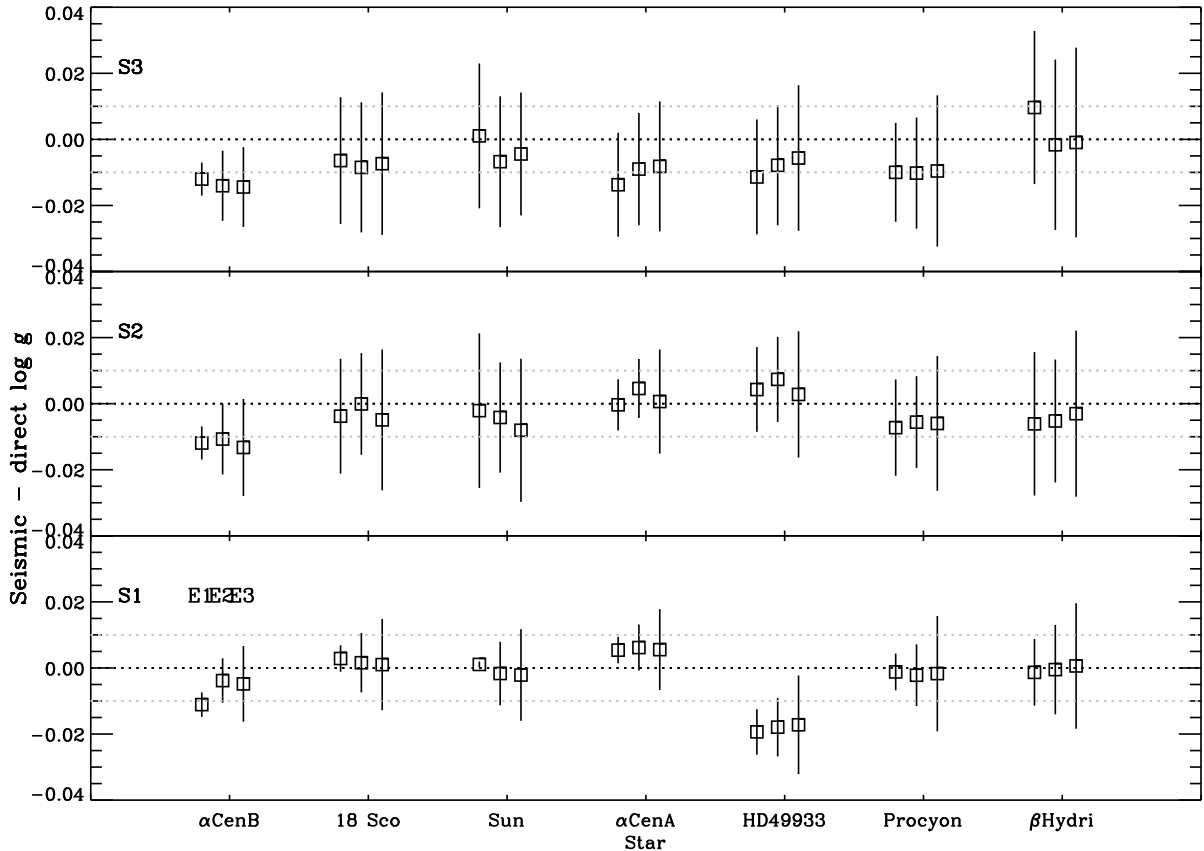}
\caption{Accuracy of method.
{\it Seismic-minus-direct} \logg\ for the seven sample stars 
while considering different subsets of input observations (different panels, 
S1, S2, S3)
and different observational errors (E1, E2, E3).  
See Sect.~\ref{sec:analysisapproach} 
for details.\label{fig:logg}}
\end{figure*}

In Figure \ref{fig:logg} we compare the grid-based seismic \logg\ with the 
direct \logg\ for the seven stars.
Each star is represented by a point on the abscissa, and the y-axis
shows (seismic - direct) value of \logg.
There are {three} panels which represent the results using the {three} 
sets of input data. 
%(these are separated by a 0.06 dex offset in the ordinate and a very slight
%shift in the abscissa just for clarity).
We also show for each star in each panel three results; in the bottom left 
corner these are marked by 'E1', 'E2', and 'E3', and they 
represent the results using the different errors in the observations.
The black dotted lines represent {\it seismic minus direct \logg\ =} 0, 
and the grey
dotted lines indicate \pmm 0.01 dex.

Figure~\ref{fig:logg} shows that for all observational sets and errors
\logg\ is generally estimated to within 0.01 dex in accuracy and with
a precision of 0.015 dex.
This result clearly shows the validity of the global seismic quantities and 
atmospheric parameters for providing accurate values of \logg.
It can also be noted %that %Other general trends that can be seen are 
that the {\it precision}
in \logg\ typically decreases as (1) the observational errors increase (from E2 -- E3),
and (2) the information content decreases (S1 -- S2 -- S3, for example).

One noticeable result from Figure~\ref{fig:logg} 
is the systematic offset in the derivation of \logg\ for
HD~49933 when we use \feh\ as input (S1).
This could be due to an incorrect \feh\ or \teff, an error in the adopted 
direct \logg\ or a shortcoming of the grid of models.
This star is known to be active and \citet{mos05} found clear evidence of spot signatures in line bisectors. More recently \citet{gar10science} and \citet{sal11} found frequency and amplitude  variations similar to those in the Sun, evidence of the presence of a stellar cycle.  
\citet{gar10science} derived an S-index of about 0.3 corresponding to a very active star, even though no detection of a magnetic field 
has yet been confirmed (P. Petit, private communication 2012).
\citet{fab12} showed that the magnetic effects in stellar models affect the determination of the metallicity
and this, in turn, will affect 
the determination of other stellar parameters such as mass. 
In their work, they considered large magnetic field strengths of several hundreds of Gauss. 

Since no evidence of such a large magnetic field has yet been found for HD~49933 
we propose another explanation. 
We consider the effect of the confirmed presence of spots on the effective temperature. 
{By considering a spot area of about 20 percent of the stellar surface we found that the real effective temperature 
(that of the non spotted surface of the star) should increase by about 300 K with respect to the non-spotted star. Adding 300 K
to \teff\ results in a seismic \logg\ that increases by $0.011$ dex for S1 (considering errors E1 and E2),
 %Such increase of Teff injected in our procedure leads to an increase of the derived log g by about 0.01 dex 
which makes the new value consistent with the direct value within error bars.
The increase of 0.011 dex corresponds to a relative increase of 3.7$\sigma$ and 1.2$\sigma$ for both error types E1 and E2.
If we add this same 300 K for cases S2 and S3, we also find an increase in \logg, but smaller (0.009 and 0.006 dex),
corresponding to a relative increase of 0.6$\sigma$ and 0.4$\sigma$, respectively.} %or E1 and E2.

%This star shows stellar activity \citep{mos05,gar10science,sal11} and 
%not including magnetic effects in stellar models will effect the 
%determination of \feh, e.g. \citet{fab12} and consequently the determination
%of other stellar parameters such as mass. The effect of magnetic fields
%on the seismic parameters has been shown to be smaller than the errors \red{cite bigot}.
%The typical spread in \logg\ for this star
%as shown in Table~\ref{tab:refstars} is about 0.02 
%dex which is also the size of the offset.  
%However, this 0.02 dex offset is still much better than any spectroscopic
%determination of \logg.
%In Table~\ref{tab:radexgmr} we summarize the derived value of \logg\ and 
%the direct value using S1 with the true observational errors (E1).

\begin{table}
\begin{center}
\caption{\logg\ derived by RadEx10 for the reference stars
using the true measurement errors. $\Delta g = \log g - \log g_{\rm direct}$.\label{tab:radexgmr}}
\begin{tabular}{lccccccccccccccccccc}
\hline\hline
Star &  \logg & $\log g_{\rm direct}$ & $\Delta g$& $\Delta g$\\   %$R$ & $M$ & $L$ & Age\\
& (dex) & (dex) & (dex) & ($\sigma$)\\ %(\rsol)&(\msol) & (L$_{\odot}$) & Gyr\\
\hline
$\alpha$ Cen B & 4.527 \pmm\ 0.004 & 4.538 & --0.011& --2.8 \\%& 0.859 \pmm\ 0.007 & 0.905 \pmm\ 0.023 &  0.52 \pmm\  0.02 &  9.4 \pmm\  2.0\\
18 Sco & 4.441 \pmm\ 0.004 & 4.438 & 0.003 & 0.8\\%1.018 \pmm\ 0.008 & 1.042 \pmm\ 0.019 &  1.07 \pmm\  0.04 &  4.8 \pmm\  0.9\\
Sun & 4.438 \pmm\ 0.001 & 4.437 & 0.001 &1.0\\%1.000 \pmm\ 0.002 & 1.000 \pmm\ 0.005 &  1.01 \pmm\  0.03 &  6.3 \pmm\  0.6\\
$\alpha$ Cen A & 4.312 \pmm\ 0.004 & 4.307 & 0.005& 1.3\\%1.223 \pmm\ 0.010 & 1.119 \pmm\ 0.024 &  1.56 \pmm\  0.08 &  7.0 \pmm\  0.9\\
HD\,49933 & 4.195 \pmm\ 0.007 & 4.214 & 0.019& 2.7\\%1.418 \pmm\ 0.022 & 1.148 \pmm\ 0.054 &  3.23 \pmm\  0.22 &  3.5 \pmm\  0.6\\
Procyon & 3.981 \pmm\ 0.006 & 3.982&--0.001&--0.2\\%2.072 \pmm\ 0.024 & 1.497 \pmm\ 0.041 &  7.08 \pmm\  0.54 &  2.1 \pmm\  0.2\\
$\beta$ Hydri & 3.957 \pmm\ 0.010 & 3.958& 0.001&0.1\\% 1.840 \pmm\ 0.045 & 1.119 \pmm\ 0.086 &  3.55 \pmm\  0.31 &  6.8 \pmm\  1.0\\

\hline
\end{tabular}
\end{center}
\end{table}

\subsection{Systematic errors in observations\label{sec:systerrors}}

For all of the calculations we have assumed that the input observations
are correct (accurate).  While this is certainly more true for brighter 
nearby targets where high SNR data can be obtained, the same cannot 
be said for fainter stars.  In particular with spectroscopic 
data, the determination of \teff\ and 
\feh\ are correlated and depend on the analysis methods used and the different
model atmospheres 
(see e.g. \citealt{cre12kep,leb12}).  
Additionally for many stars a photometric temperature may be the 
only available one and while these estimates are very good, systematic
errors are still unavoidable \citep{cas10,boy12}, in particular due to 
unknown reddenning.
Larger photometric errors also lead to larger errors on the temperatures.
%and
%systematics in temperature scales \citep{cas10,boy12}.
This is not only a problem for fainter stars.  
For example, for $\beta$ Hydri we found two determinations of 
\teff\ --- an interferometric one and a spectroscopic one.  
For the spectroscopic \teff, the corresponding fitted metallicity from
the atmospheric analysis will correlate with it. %The metallicity derived from spectroscopic data are inherently 
%dependent on the spectroscopic \teff.
%Spectroscopic determinations of \teff\ will also necessarily influence 
%on the adopted \teff. %the derived value of \feh.
%The spectroscopic \teff\ also has a different \feh\ than that cited by 
%the authors who obtained the interferometric \teff.

\begin{figure}
\includegraphics[width=0.49\textwidth]{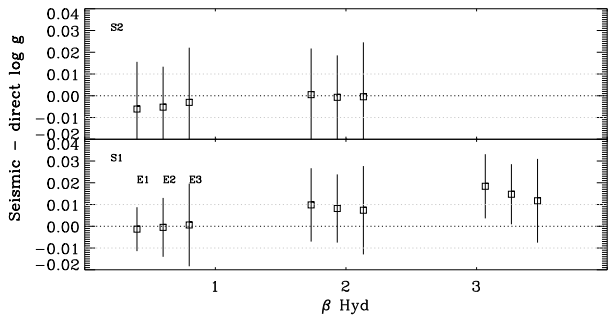}
\caption{{\it Seismic-minus-direct} values of \logg\ for $\beta$ Hydri when
the observations with systematic errors included for cases 2 and 3 
are analysed by the 
seismic method. See Sect.~\ref{sec:systerrors}.{\label{fig:syst}}}
\end{figure}

To study the effect of {\it systematic errors} in the observations, we 
repeated our analysis for $\beta$ Hydri while using three sets of input
data that change only in \teff\ and \feh.  
The first set (1) uses the interferometric value of \teff\ from \citet{nor07} 
and their \feh\ 
(5872, --0.10), which we consider as the most correct, the second set (2) uses (5964, --0.10), 
and the third set (3) uses (5964, --0.03) as given by \citet{das06}.
The results for S1 and S2 are shown in Figure~\ref{fig:syst}. 
%(the 
%results for S4 are very similar to those for S2).
The lower panel shows that for a systematic error in both \teff\ and \feh\ 
the accuracy decreases.  The top panel shows that when we only consider
the \teff\ information we get a smaller increase in the offset than when 
we consider both \teff\ and \feh. 
One way of interpreting this result is by considering that in S2 there is 
much more weight assigned to the seismic data than the atmospheric data, and 
so an incorrect atmospheric parameter should not influence the final
result as much as in case S1 where the atmospheric parameters have more weight.
In the latter case, an incorrect \teff\ with the correct \feh\ will necessarily
shift the mass either up or down (in terms of the H-R diagram), 
and result in a more displaced \logg.
However, in both cases, we see that the offset does not exceed 0.02 dex.
For determining a seismic radius and mass, 
however, a systematic error in the input observations has a much more profound
effect on the offset.  In this case biases on the order of up to 6\% and 20\% 
in radius and mass can be obtained \citep{cre12proc}. 
It must also be noted that a systematic error in the atmospheric parameters
is going to have a much larger negative 
effect when we use only the {\it global} seismic 
quantities instead of performing a 
detailed seismic analysis with individual frequencies, where the latter 
have much more weight in the fitting process.
%is less susceptible to the systematic errors.

%\subsection{discussion of results}
%    -log g MI vs MD

%    -log g effect of errors

%    -R

%    -M

%    -Age (?)

%    -including FeH

%    -using Dnu vs numax

%    -no atmospheric constraints

%\subsection{Stellar properties of a sample of solar-type stars}

%(some Kepler targets, some ground-based....)  table of logg, R... 
%show comparison with independent measures if available.

\subsection{Seismic determination of \logg\ from the global seismic quantities for the 
reference stars}
We summarize \logg\ for the sample stars in 
Table~\ref{tab:radexgmr}
derived
by RadEx10 using \mlsep, \numax, \teff, and \feh, and the true observational
errors.  
We highlight the excellent agreement between our seismically determined
parameters, and those obtained by direct mass and radius estimates.  
\logg\ is matched to within $\sim$0.01 dex, and 0.02 dex for HD\,49933.
%, and if we
%consider the more relaxed observational errors (Fig.~\ref{fig:logg})
%then the accuracy is to within a 1$\sigma$ uncertainty.
%, and 
%in only 
%two cases ($\alpha$ Cen B and 18 Sco), we find that the mass is determined 
%with a difference of just over 1$\sigma$, with all of the other radius and 
%mass determinations accurate to within the quoted error bars 
%(compare Tables~\ref{tab:reflogg} and \ref{tab:radexgmr}, and see 
%Fig.~\ref{fig:radmasss0}).
We must also comment on the very small uncertainties given in 
Table~\ref{tab:radexgmr}; these results were obtained using the true 
(very small) observational
errors given in Table~\ref{tab:refstars} and typically one would not expect
to obtain such small errors for fainter stars. 
As can be seen, the results using the relaxed observational errors 
give more reasonable parameter uncertainties, i.e. 0.010 -- 0.015 dex,
see Fig.~\ref{fig:logg}, while still matching the direct \logg\ to within
1$\sigma$.
%Our deduced ages can be incorrect (e.g. 6.3 Gyr for the Sun), but with
%the use of simplified physics and especially ignoring diffusion effects, the
%ages are severely hampered.  Here it is especially important to explore
%the effects of convective overshoot, when a convective core exists 
%($M>1.2$ \msol).

\section{Determination of \logg\ for an extended list of stars 
\label{sec:extlist}}
We apply our grid-based method to an extended list of stars with 
measured global seismic quantities and atmospheric parameters.
Table~\ref{tab:extlist} 
lists the star name along with the other 
measured parameters that are used as the input for our method.
The first part of the table comprises primarily bright stars whose
oscillation properties have been measured either 
from ground-based or spaced-based instrumentation
(see references given in the table).
For most of these stars no errors are cited for \mlsep\ and
\numax.
The second part of the 
table lists a set of 22 solar-type stars observed by the \kep\ spacecraft
and studied in \citet{mat12}.  We have taken the seismic and atmospheric
data directly from this paper.
To conduct a homogenous analysis of these stars we adopted a 1.1 \mhz\ error
on \mlsep\ for all of the stars, and a 5\% error on \numax, typical of what
has been found for the large sample of \kep\ stars (see \citealt{hub11} and Sect.~\ref{sec:kepobs}).
%We note that the errors on the seismic quantities from the \kep\ data 
%have been increased at least two-fold; our justification for doing this is
%because \logg\ has been determined using detailed seismic analyses from 
%individual frequencies and such precisions are not typical.

\begin{table*}
\begin{center}
\caption{Measured seismic and atmospheric properties for an extended
list of Sun-like oscillators\label{tab:extlist}}
\begin{tabular}{lccccccccc}
\hline\hline
Star Name & \mlsep\ & \numax & \teff& \feh\\
& (\mhz)&(\mhz)&(K)&(dex)\\
\hline
HD~  10700& 169.0$^a$& 4500$^a$& 5383 \pmm\    47$^a$&-0.10 \pmm\  0.07\\
HD~  17051& 120.0$^b$& 2700$^b$& 6080 \pmm\    80& 0.15 \pmm\  0.07\\
HD~  23249&  43.8$^c$&  700$^c$& 4986 \pmm\    57$^d$& 0.15 \pmm\  0.07\\
HD~ 49385& 55.8$^s$& 1013$^s$&6095 \pmm\ 65$^s$ & +0.09 \pmm\ 0.05$^s$\\
HD~ 52265&98.3 \pmm\ 0.1$^l$&1800$^l$ &6100 \pmm\ 60$^l$&0.19 \pmm\ 0.05$^l$\\
HD~  61421&  55.0& 1000& 6494 \pmm\    48& 0.01 \pmm\  0.07\\
HD~  63077&  97.0$^e$& 2050$^e$& 5710 \pmm\    80&-0.86 \pmm\  0.07\\
HD~ 102870&  72.1$^f$& 1400$^f$& 6012 \pmm\    64$^g$& 0.12 \pmm\  0.07\\
HD~ 121370&  39.9$^h$&  750$^h$& 6028 \pmm\    47$^d$& 0.24 \pmm\  0.07\\
HD~ 139211&  85.0$^e$& 1800$^e$& 6200 \pmm\    80&-0.04 \pmm\  0.07\\
HD~ 160691&  90.0$^i$& 2000$^i$& 5665 \pmm\    80& 0.32 \pmm\  0.07\\
HD~ 165341& 161.7$^j$& 4500$^j$& 5300 \pmm\    80& 0.12 \pmm\  0.07\\
HD~ 170987&55.5 \pmm\ 0.8$^m$&1000$^m$&6540 \pmm\ 100$^m$&-0.15 \pmm\ 0.06$^m$\\
%HD~ 175726&97.2 \pmm\ 0.5$^o$&2000$^o$ &6070 \pmm\ 45$^o$&-0.07 \pmm\ 0.03$^o$\\
HD~ 181420&75.0$^n$&1500 \pmm\ 300$^n$&6580 \pmm\ 105$^o$&0.00 \pmm\ 0.06$^o$\\
HD~ 181906&87.5 \pmm\ 2.6$^p$&1912 \pmm\ 47$^p$&6300 \pmm\ 150$^p$&-0.11 \pmm\ 0.14$^p$\\
HD~ 186408&103.1$^q$&2150$^q$&5825 \pmm\ 50$^q$&0.01 \pmm\ 0.03$^q$\\
HD~ 186427&117.2$^q$&2550$^q$&5750 \pmm\ 50$^q$&0.05 \pmm\ 0.02$^q$\\
HD~ 185395&84.0$^r$&2000$^r$&6745 \pmm\ 150$^r$&-0.04$^r$\\
HD~ 203608& 120.4$^k$& 2600$^k$& 5990 \pmm\    80&-0.74 \pmm\  0.07\\
HD~ 210302&  89.5$^e$& 1950$^e$& 6235 \pmm\    80& 0.01 \pmm\  0.07\\
KIC~ 3632418& 60.63 \pmm\   0.37& 1110 \pmm\    20& 6150 \pmm\    70&-0.19 \pmm\  0.07\\
KIC~ 3656476& 93.70 \pmm\   0.22& 1940 \pmm\    25& 5700 \pmm\    70& 0.32 \pmm\  0.07\\
KIC~ 4914923& 88.61 \pmm\   0.32& 1835 \pmm\    60& 5840 \pmm\    70& 0.14 \pmm\  0.07\\
KIC~ 5184732& 95.53 \pmm\   0.26& 2070 \pmm\    20& 5825 \pmm\    70& 0.39 \pmm\  0.07\\
KIC~ 5512589& 68.52 \pmm\   0.33& 1240 \pmm\    25& 5710 \pmm\    70& 0.04 \pmm\  0.07\\
KIC~ 6106415&103.82 \pmm\   0.29& 2285 \pmm\    20& 5950 \pmm\    70&-0.11 \pmm\  0.07\\
KIC~ 6116048&100.14 \pmm\   0.22& 2120 \pmm\    20& 5895 \pmm\    70&-0.26 \pmm\  0.07\\
KIC~ 6603624&110.28 \pmm\   0.25& 2405 \pmm\    50& 5600 \pmm\    70& 0.26 \pmm\  0.07\\
KIC~ 6933899& 72.15 \pmm\   0.25& 1370 \pmm\    30& 5830 \pmm\    70&-0.01 \pmm\  0.07\\
KIC~ 7680114& 85.13 \pmm\   0.14& 1660 \pmm\    25& 5815 \pmm\    70& 0.10 \pmm\  0.07\\
KIC~ 7976303& 50.95 \pmm\   0.37&  910 \pmm\    25& 6050 \pmm\    70&-0.52 \pmm\  0.07\\
KIC~ 8006161&148.21 \pmm\   0.19& 3545 \pmm\   140& 5340 \pmm\    70& 0.38 \pmm\  0.07\\
KIC~ 8228742& 63.15 \pmm\   0.32& 1160 \pmm\    40& 6000 \pmm\    70&-0.15 \pmm\  0.07\\
KIC~ 8379927&120.86 \pmm\   0.43& 2880 \pmm\    65& 5960 \pmm\   125&-0.30 \pmm\  0.07\\
KIC~ 8760414&116.24 \pmm\   0.56& 2510 \pmm\    95& 5765 \pmm\    70&-1.19 \pmm\  0.07\\
KIC~10018963& 55.99 \pmm\   0.35&  985 \pmm\    10& 6300 \pmm\    65&-0.47 \pmm\  0.50\\
KIC~10516096& 84.15 \pmm\   0.36& 1710 \pmm\    15& 5900 \pmm\    70&-0.10 \pmm\  0.07\\
KIC~10963065&103.61 \pmm\   0.41& 2160 \pmm\    35& 6015 \pmm\    70&-0.21 \pmm\  0.07\\
KIC~11244118& 71.68 \pmm\   0.16& 1405 \pmm\    20& 5705 \pmm\    70& 0.34 \pmm\  0.07\\
KIC~11713510& 69.22 \pmm\   0.20& 1235 \pmm\    15& 5930 \pmm\    52& ... \\
KIC~12009504& 88.10 \pmm\   0.42& 1825 \pmm\    20& 6060 \pmm\    70&-0.09 \pmm\  0.07\\
KIC~12258514& 74.75 \pmm\   0.23& 1475 \pmm\    30& 5950 \pmm\    70& 0.02 \pmm\  0.07\\
\hline\hline
\end{tabular}
\end{center}
References:
All HD star measurements without references are taken from \citet{bru10},
$^a$\citet{tei09},   %tau cet, 10700  both
$^b$\citet{vau08},       %i Hor, 17051 seismic
$^c$\citet{bc03}, $^d$\citet{the05},  %alpha men, 43834
$^e$\citet{bru10}, %63077 171 pup
$^f$\citet{car05bvir}, $^g$\citet{nor09}, %beta Vir 102870 
%^a$\citet{the05}, 
$^h$\citet{car05eboo}, %eta boo 121370
$^i$\citet{bou05},  %muara 160691
$^j$\citet{ce06},  %70 oph a 165341
$^k$\citet{mos08}, %gamma Pav 203608
%$^j$\citet{bru10}.\\ %210302  private communication
$^l$\citet{ballot11}, %hd 52265
$^m$\citet{mat170987},
$^n$\citet{barban09}, %% hd 181420
$^o$\citet{bruntt09}, %% hd 181420
$^p$\citet{garcia09},  %% 181906
$^q$\citet{met16cyg}, %%hd 186408, 186427
$^r$\citet{guzik11}, %% hd 185395  13 cyg
$^s$\citet{deh10}. %% hd 49385
When \mlsep\ is not explicitly given it is calculated from the highest amplitude $l=0$ modes.
Values for the KIC stars are taken from \citet{mat12}. 
\end{table*}

%\red{include??}
%Over the past decade or so, many ground-based observational campaigns
%have been dedicated to detecting solar-like oscillations in stars.  
%In this section we determine the stellar properties of these stars using
%the method described in Sect.~\ref{}.  

\begin{table}
\begin{center}
\caption{Derived seismic \logg\ for an extended
list of Sun-like oscillators.
\logg\ and \logg$_{\rm no [Fe/H]}$ refer to using 
\{\mlsep,\numax,\teff,\feh\} and \{\mlsep,\numax,\teff\}
as the constraints in the analysis.  
For a homogenous analysis we adopted 1.1 \mhz\ and 5\% as 
the errors on \mlsep\ and \numax. 
We list 2$\sigma$ uncertainties for \logg. % and 1$\sigma$ for the rest.
 \label{tab:derivedextlist}}
\begin{tabular}{lccccccccc}
\hline\hline
Star Name & \logg & \logg$_{\rm no [Fe/H]}$ \\%& $R$ & $M$ & $L$ & $t$\\
& (dex) & (dex) \\%& (\rsol) & (\msol) & (\lsol) & (Gyr)\\
\hline
HD~  10700&4.55 \pmm\ 0.02&4.57 \pmm\ 0.02\\%&0.82 \pmm\ 0.02&0.87 \pmm\ 0.05& 0.5 \pmm\  0.0& 7.3 \pmm\  3.7\\
HD~  17051&4.40 \pmm\ 0.02&4.39 \pmm\ 0.03\\%&1.15 \pmm\ 0.02&1.20 \pmm\ 0.06& 1.6 \pmm\  0.1& 1.9 \pmm\  1.3\\
HD~  23249&3.81 \pmm\ 0.02&3.78 \pmm\ 0.04\\%&2.03 \pmm\ 0.04&0.96 \pmm\ 0.06& 2.3 \pmm\  0.2&14.3 \pmm\  2.6\\
HD~ 49385&3.98 \pmm\ 0.02 & 3.97 \pmm\ 0.04\\
HD~ 52265&4.28 \pmm\ 0.02&4.24 \pmm\ 0.02\\
HD~  61421&3.98 \pmm\ 0.02&3.97 \pmm\ 0.03\\%&2.10 \pmm\ 0.04&1.53 \pmm\ 0.06& 7.0 \pmm\  0.4& 2.1 \pmm\  0.2\\
HD~  63077&4.22 \pmm\ 0.02&4.25 \pmm\ 0.03\\%&1.15 \pmm\ 0.01&0.80 \pmm\ 0.01& 1.4 \pmm\  0.0&15.5 \pmm\  0.4\\
HD~ 102870&4.11 \pmm\ 0.02&4.10 \pmm\ 0.04\\%&1.66 \pmm\ 0.04&1.30 \pmm\ 0.07& 3.2 \pmm\  0.2& 4.4 \pmm\  1.0\\
HD~ 121370&3.82 \pmm\ 0.03&3.82 \pmm\ 0.03\\%&2.65 \pmm\ 0.06&1.67 \pmm\ 0.03& 8.4 \pmm\  0.6& 2.4 \pmm\  0.1\\
HD~ 139211&4.20 \pmm\ 0.02&4.21 \pmm\ 0.02\\%&1.45 \pmm\ 0.03&1.21 \pmm\ 0.06& 2.8 \pmm\  0.2& 4.0 \pmm\  0.8\\
HD~ 160691&4.22 \pmm\ 0.02&4.21 \pmm\ 0.02\\%&1.33 \pmm\ 0.03&1.06 \pmm\ 0.05& 1.7 \pmm\  0.2&11.0 \pmm\  2.3\\
HD~ 165341&4.54 \pmm\ 0.02&4.54 \pmm\ 0.02\\%&0.87 \pmm\ 0.02&0.95 \pmm\ 0.05& 0.5 \pmm\  0.1& 6.5 \pmm\  3.6\\
HD~ 170987&3.98 \pmm\ 0.02&3.98 \pmm\ 0.03\\
HD~ 181420&4.15 \pmm\ 0.02&4.15 \pmm\ 0.03\\
HD~ 181906&4.22 \pmm\ 0.02&4.23 \pmm\ 0.02\\
%HD~ 175726&4.26 \pmm\ 0.02&4.26 \pmm\ 0.03\\
HD~ 186408&4.28 \pmm\ 0.02&4.28 \pmm\ 0.03\\
HD~ 186427&4.36 \pmm\ 0.02&4.35 \pmm\ 0.03\\
HD~ 185395&4.22 \pmm\ 0.02&4.23 \pmm\ 0.02\\
HD~ 203608&4.35 \pmm\ 0.02&4.37 \pmm\ 0.04\\%&1.02 \pmm\ 0.02&0.84 \pmm\ 0.03& 1.2 \pmm\  0.1&11.1 \pmm\  2.4\\
HD~ 210302&4.23 \pmm\ 0.02&4.24 \pmm\ 0.03\\%&1.41 \pmm\ 0.02&1.24 \pmm\ 0.06& 2.7 \pmm\  0.2& 3.4 \pmm\  0.8\\
KIC~ 3632418&4.00 \pmm\ 0.03&4.01 \pmm\ 0.04\\%&1.82 \pmm\ 0.07&1.23 \pmm\ 0.11& 4.3 \pmm\  0.5& 4.3 \pmm\  1.3\\
KIC~ 3656476&4.23 \pmm\ 0.02&4.23 \pmm\ 0.03\\%&1.30 \pmm\ 0.02&1.06 \pmm\ 0.04& 1.6 \pmm\  0.1&10.8 \pmm\  1.8\\
KIC~ 4914923&4.21 \pmm\ 0.02&4.21 \pmm\ 0.03\\%&1.38 \pmm\ 0.03&1.14 \pmm\ 0.06& 2.0 \pmm\  0.2& 7.4 \pmm\  1.8\\
KIC~ 5184732&4.26 \pmm\ 0.02&4.25 \pmm\ 0.02\\%&1.31 \pmm\ 0.02&1.12 \pmm\ 0.04& 1.8 \pmm\  0.1& 7.8 \pmm\  1.6\\
KIC~ 5512589&4.05 \pmm\ 0.02&4.04 \pmm\ 0.04\\%&1.61 \pmm\ 0.04&1.07 \pmm\ 0.06& 2.5 \pmm\  0.2& 9.4 \pmm\  1.4\\
KIC~ 6106415&4.29 \pmm\ 0.02&4.30 \pmm\ 0.03\\%&1.19 \pmm\ 0.02&1.01 \pmm\ 0.06& 1.6 \pmm\  0.1& 8.2 \pmm\  1.8\\
KIC~ 6116048&4.25 \pmm\ 0.02&4.27 \pmm\ 0.03\\%&1.17 \pmm\ 0.03&0.90 \pmm\ 0.05& 1.5 \pmm\  0.1&11.9 \pmm\  2.4\\
KIC~ 6603624&4.32 \pmm\ 0.02&4.32 \pmm\ 0.02\\%&1.14 \pmm\ 0.02&1.00 \pmm\ 0.03& 1.2 \pmm\  0.1&12.7 \pmm\  2.3\\
KIC~ 6933899&4.09 \pmm\ 0.02&4.09 \pmm\ 0.04\\%&1.56 \pmm\ 0.05&1.09 \pmm\ 0.08& 2.5 \pmm\  0.2& 8.2 \pmm\  1.9\\
KIC~ 7680114&4.18 \pmm\ 0.02&4.17 \pmm\ 0.04\\%&1.40 \pmm\ 0.03&1.09 \pmm\ 0.07& 2.0 \pmm\  0.2& 8.8 \pmm\  1.9\\
KIC~ 7976303&3.89 \pmm\ 0.02&3.92 \pmm\ 0.04\\%&1.92 \pmm\ 0.03&1.03 \pmm\ 0.02& 4.4 \pmm\  0.3& 6.2 \pmm\  0.6\\
KIC~ 8006161&4.48 \pmm\ 0.02&4.48 \pmm\ 0.02\\%&0.92 \pmm\ 0.01&0.93 \pmm\ 0.03& 0.6 \pmm\  0.0&12.2 \pmm\  2.7\\
KIC~ 8228742&4.02 \pmm\ 0.03&4.02 \pmm\ 0.04\\%&1.73 \pmm\ 0.06&1.15 \pmm\ 0.11& 3.5 \pmm\  0.4& 5.9 \pmm\  1.6\\
KIC~ 8379927&4.36 \pmm\ 0.02&4.40 \pmm\ 0.02\\%&1.04 \pmm\ 0.03&0.92 \pmm\ 0.06& 1.3 \pmm\  0.2& 8.7 \pmm\  3.1\\
KIC~ 8760414&4.31 \pmm\ 0.02&4.35 \pmm\ 0.03\\%&1.01 \pmm\ 0.01&0.76 \pmm\ 0.02& 1.0 \pmm\  0.1&18.2 \pmm\  2.1\\
KIC~10018963&3.96 \pmm\ 0.03&3.96 \pmm\ 0.03\\%&1.95 \pmm\ 0.10&1.28 \pmm\ 0.15& 5.4 \pmm\  0.7& 3.6 \pmm\  1.2\\
KIC~10516096&4.17 \pmm\ 0.02&4.18 \pmm\ 0.03\\%&1.39 \pmm\ 0.04&1.04 \pmm\ 0.08& 2.1 \pmm\  0.2& 8.6 \pmm\  2.2\\
KIC~10963065&4.28 \pmm\ 0.02&4.29 \pmm\ 0.04\\%&1.18 \pmm\ 0.03&0.96 \pmm\ 0.07& 1.6 \pmm\  0.1& 8.7 \pmm\  2.1\\
KIC~11244118&4.09 \pmm\ 0.02&4.08 \pmm\ 0.03\\%&1.60 \pmm\ 0.04&1.15 \pmm\ 0.07& 2.4 \pmm\  0.2& 8.6 \pmm\  1.8\\
KIC~11713510&4.05 \pmm\ 0.04&4.05 \pmm\ 0.03\\%&1.58 \pmm\ 0.08&1.03 \pmm\ 0.14& 2.8 \pmm\  0.3& 8.4 \pmm\  2.3\\
KIC~12009504&4.21 \pmm\ 0.02&4.21 \pmm\ 0.03\\%&1.37 \pmm\ 0.03&1.10 \pmm\ 0.07& 2.3 \pmm\  0.2& 6.2 \pmm\  1.6\\
KIC~12258514&4.12 \pmm\ 0.02&4.12 \pmm\ 0.03\\%&1.57 \pmm\ 0.04&1.19 \pmm\ 0.08& 2.8 \pmm\  0.2& 5.8 \pmm\  1.5\\

\hline\hline
\end{tabular}
\end{center}
\end{table}

Table~\ref{tab:derivedextlist} lists the derived value of \logg\ and 
2$\sigma$ uncertainties (to allow for round-up error)
for each
of the stars using RadEx10.  We show two values of \logg; the first
is obtained by using the four input constraints \{\mlsep,\numax,\teff,\feh\}
and the second $\log g_{\rm no [Fe/H]}$ is obtained by omitting \feh\ from
the analysis.
%The other stellar properties are obtained using the four input constraints.
%We list 2$\sigma$ uncertainties for \logg\ and
%1$\sigma$ uncertainties for $R$, $M$, $L$, and age $t$.

\begin{figure}
\includegraphics[width=0.48\textwidth]{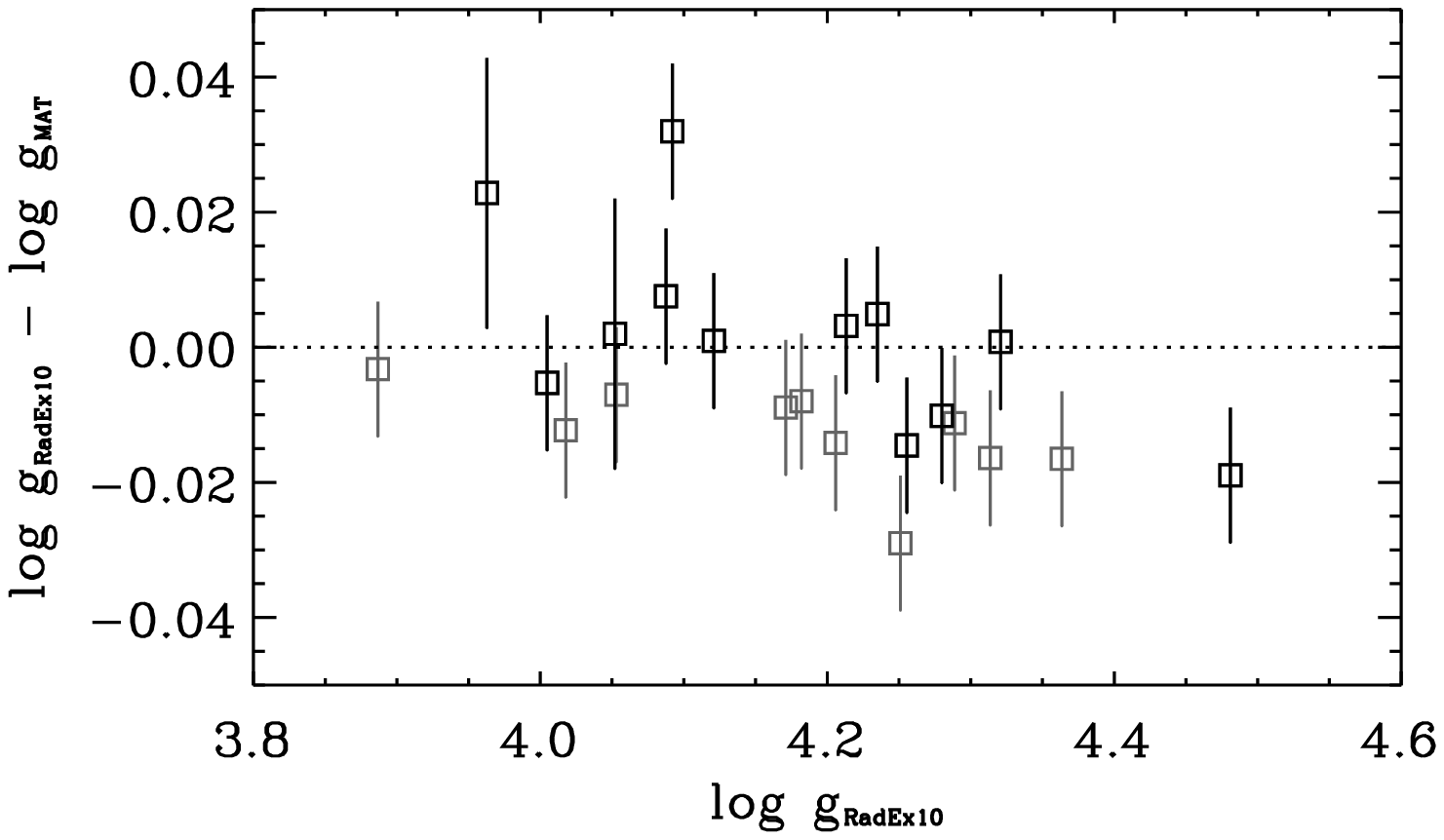}
\includegraphics[width=0.48\textwidth]{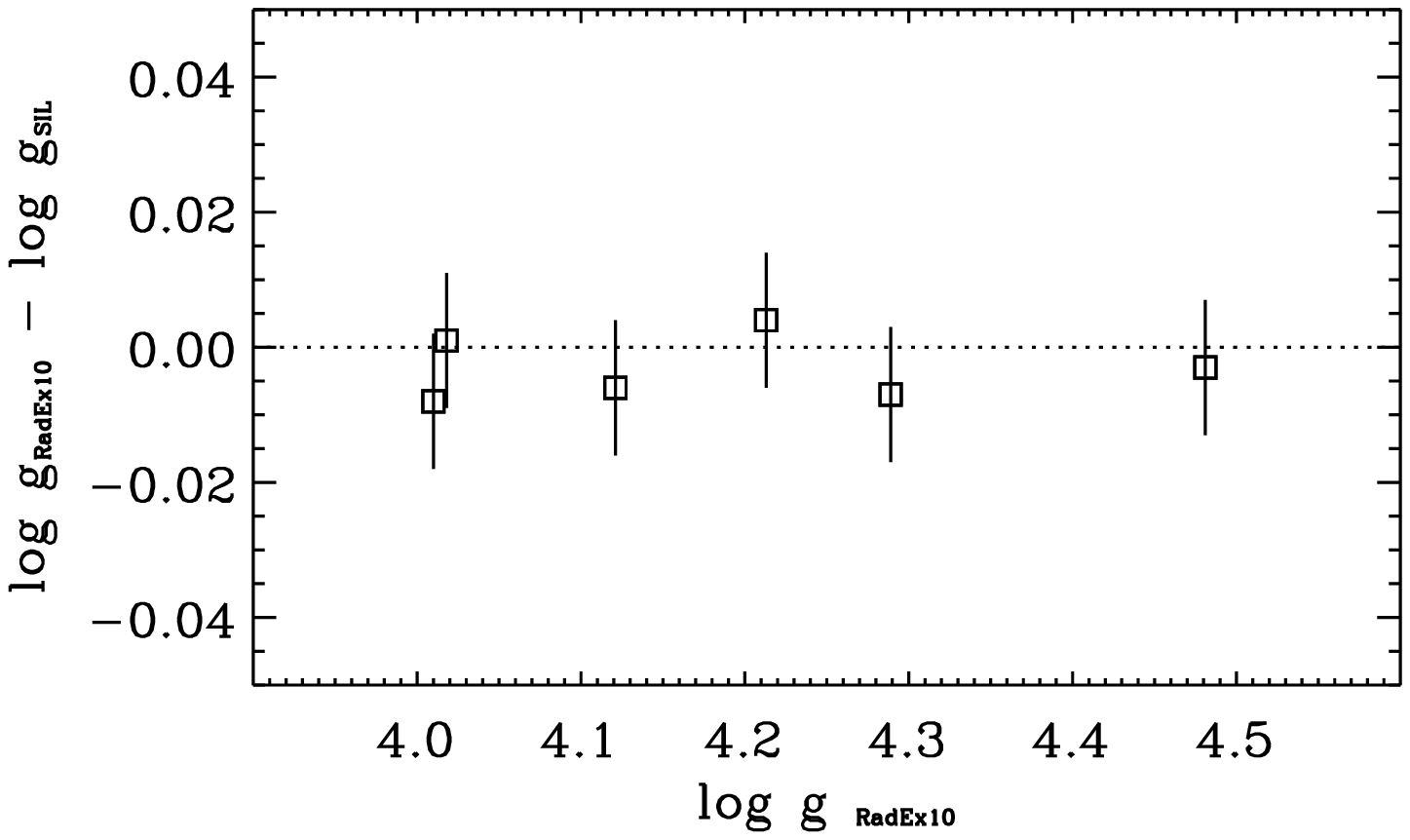}
\caption{Comparison of \logg\ derived by RadEx10 with 
(top) 
detailed seismic analyses \citep{mat12}
for 22 Sun-like stars observed by \kep, 
and (bottom) six common stars from \citet{sil12}. \label{fig:kicradex10logg}}
\end{figure}

Figure~\ref{fig:kicradex10logg} top panel 
compares the derived values of \logg\ for the \kep\ stars 
with those determined
using the individual oscillation frequencies, as given by \citet{mat12},
with our 1$\sigma$ error bars overplotted.
We see that the grid-based method provides \logg\ consistent with those 
derived from a detailed asteroseismic analysis, although a very small trend can
be seen.
For some of the stars they obtain a fitted initial He abundance 
significantly below the accepted primordial value,
suggesting that the 
%of $Y_0= 0.22$, at the edge of their searched parameter space in metallicity.
%This could indicate that the 
corresponding fitted mass and radius may be 
slightly biased (grey squares).   
%not 
%the optimal models.
%, i.e.
%sub-Big Bang values, and at the edge of their searched parameter space.  
%The corresponding fitted mass and radius may then be slightly skewed.
%If we omit these two extreme values 
%(the \citet{mat12} values of He abundance was only $Y_0 = 0.22$, i.e.
%sub-Big Bang value which renders suspect the other fitted parameters), 
%and the point with the largest error bar (this had a poor $\chi^2$ fit
%to the oscillation frequencies) 
If we omit these stars, 
then we fit a slope of {-0.03 \pmm\ 0.02} to the 
difference between their and our \logg\ values.
%, implying that the trend is not statistically significant.
%We note that their results were obtained by using individual 
%frequencies as the observations and not the global seismic quantities
%with scaling relations, and as 
{\citet{whi11} and \citet {mos12b} point out that a discrepancy 
may exist between the two different theoretical values of \mlsep\ 
(using scaling relations or from individual frequencies), 
and the two different approaches could also be responsible for this trend.
%
%and 
%this in turn can result in slightly different stellar properties.
We see, however, that our values agree generally to within 0.01 dex.}
\citet{sil12} have analysed 6 of these stars  and the lower panel
of Fig.~\ref{fig:kicradex10logg} shows a comparison between 
their \logg\ values with ours. 
{Fitting the differences between our results, we obtain a 
slope of 0.001 \pmm\ 0.025
with a systematic offset of -0.003 \pmm\ 0.005, indicating 
no systematic trends.}

%A comparison with the derived values from \citet{mat12} reveals good consistency
%for the radius, slightly poorer consistency for the mass values, and 
%the age values that we list are systematically overestimated.  
%This is partially due to the input physics used in the grid, where diffusion
%is not included, as well as discrepancies with mass values, which therefore
%lead to corresponding offsets in the derived age, even if the evolution stage is
%the same.

%\pagebreak

%\pagebreak
\section{Precision in \logg\ for a large sample of \kep\ stars of classes IV/V
\label{sec:kepstars}}
%Stellar properties of an extended sample of bright stars}

Our primary objective was to test the accuracy of a seismic \logg\ 
by using bright nearby targets that have independent mass and radius 
measurements.
%determine how accurately 
%we can derive \logg\ from the mean seismic quantities for some well known
%stars.  
We showed in Sect.~\ref{sec:seismicdirect} (see Fig.~\ref{fig:logg}) 
that our accuracy should be on a level of 
0.01 dex with a precision of $\sim$0.02 dex %.within 1$\sigma$ of the derived uncertainty, 
using 
the set 
%\{\mlsep,\numax,\teff,\feh\} or 
\{\mlsep,\numax,\teff\}
for the small sample 
of stars covering the range $3.9 < \log g < 4.6$.
In this section we 
investigate the {\it precision} in \logg\ for a sample of 403 V/IV stars 
(\logg\ $>$ 3.75) observed
with the \kep\ spacecraft by employing the same analysis methods.
In particular we pay attention to systematic errors by 
1) using different sets of 
observational constraints,
2) comparing
results using two different methods which incorporate different stellar 
evolution codes and physics, and 3) 
we also show the distribution of errors as a function of magnitude 
and \logg\ and summarize the uncertainties and systematics as 
a function of magnitude.
% for an extended sample of 588 stars with \logg $> 3.0$.

\subsection{Observations\label{sec:kepobs}}

%The \kep\ data was observed during the survey period of the \kep\ mission.
During the first 9 months of the \kep\ mission
targets to be monitored with a 1 minute cadence during 1 month each 
were selected by the KASC\footnote{\kep\ Asteroseismic Science Consortium, see
\url{http://astro.phys.au.dk/KASC/}}.
These stars were chosen based on information available in the 
{\it Kepler Input Catalog}, KIC, \citep{bro11kic} 
and were expected to exhibit solar-like 
oscillations.
%Of the XX targets \red{how many?} monitored and analysed, 
A total of 588 stars with values of \logg\ between 3.0 and 4.5 dex 
were analysed \citep{garcia11kep} and 
had their global
seismic quantities determined \citep{hub11, ver11, cha11science}.
%\red{are the dnu and numax published anywhere?}
{In this paper we concentrate on a subset of 403 less-evolved stars with 
\logg\  values between 3.75 and 4.50 dex derived from RadEx10, 
the range for which we have 
validated our method.}
%To illustrate the cumulative distribution of errors for the full sample, 

\begin{figure}
\includegraphics[width=0.48\textwidth]{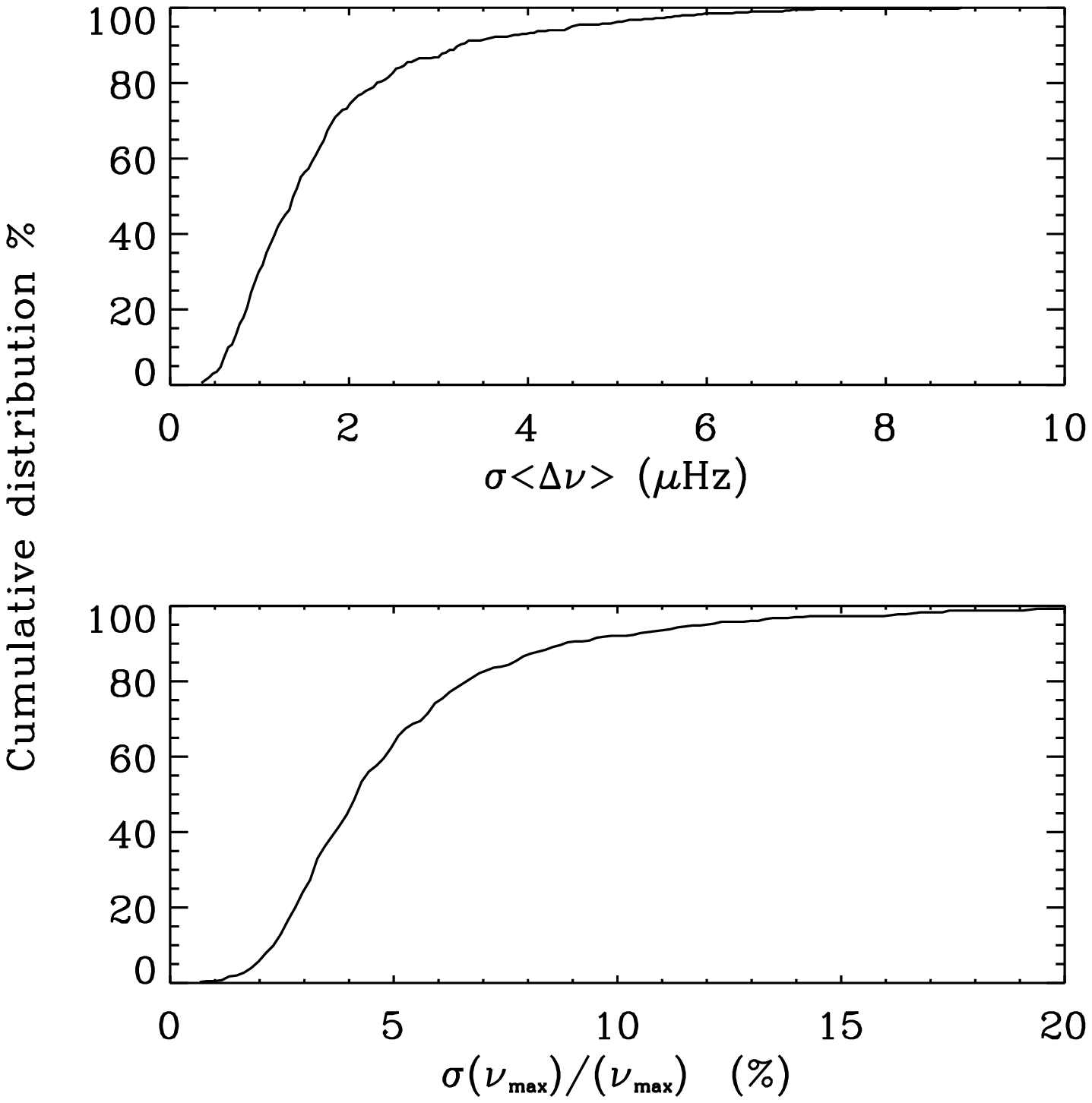}
\caption{
Cumulative distributions of the errors in \mlsep\ (top) and 
\numax\ (bottom) for 403 \kep\ stars with derived \logg\
between 3.75 and 4.5 dex. Data taken from \citet{hub11}. %The units are \mhz\ and \%, respectively. 
\label{fig:cd}}
\end{figure}

The global seismic quantities have been determined using the SYD
pipeline as described by \citet{sydpipeline}, which
 uses the reference values of 
\mlsep$_{\odot}$ = 135.1 \mhz\ and \numax$_{\odot}$ = 3,090 \mhz.
To avoid systematic errors  
(Chaplin. et al. in prep.), we adopted these same values in our grid.
The uncertainties on the seismic quantities include a contribution from
the scatter between different analysis pipelines (\citealt{ver11}, Chaplin
et al. in prep.).
Figure~\ref{fig:cd} shows the cumulative distribution for the errors
in the seismic observations, \mlsep\ (top) and \numax\ (bottom)  
in units of \mhz\ and \%, respectively, for the 403 stars.
Our choice for absolute and relative errors is for consistency with 
units used in the recent literature.  
We show the cumulative distribution of the errors 
in order to see the typical errors
for 50\% and 80\% of the sample, which justifies the 
errors that we used in Sect.~\ref{sec:analysisapproach}.
These are less
than 1.1/2.0 \mhz\ for \mlsep, and less than 5\%/8\% for \numax, respectively.

%and in 
%Table~\ref{tab:kepdata} we give a short list of these targets.

The \teff\ derived by the KIC have been shown not to be accurate on a 
star-to-star basis \citep{mol11}, 
and so the ground-based \kep\ support photometry \citep{bro11kic}
was re-analysed by \citet{pin12}
to determine more accurate \teff\ 
for the ensemble of \kep\ stars.
These are the temperatures that we adopt for our first analysis and 
we refer to them as \teff$_{\rm Pin}$.
In their work they consider a mean \feh\ = -- 0.20 \pmm\ 0.30 
for all of the stars.
%and these are the temperatures that we adopt here.
\citet{sil12} adapt an infra-red flux method presented by \citet{cas10} for 
determining stellar temperatures.  In their work they apply this method
to the large ensemble of \kep\ stars to provide an alternative determination
of \teff.   We also adopt their \teff\ determinations in order to study the 
effect of biases in temperature estimates on the derived value of \logg.
We refer to these temperature estimates as \teff$_{\rm irfm}$.
%originally presented by
%\citet{cas10} for determining stellar properties of a sample of \kep\ stars.
%These have also been made available to the community and 
Support spectroscopic data have also been obtained for 93 of the stars and
the metallicities are presented in
\citet{bru12}.
We further include these data to study the effects of possible biases
arising from lack/inclusion of metallicity information.

%AlthoughWhile we note that the precision in \logg\ does not change significantly
%when excluding \feh, %, although 
%%We also compare the results using the Bruntt et al. metallicities and the 
%%mean \feh\ = -0.20 metallicities.
%%%However, due to the much smaller number of targets, we 
%%%do not include those metallicities here.
%we have seen in Sect.~\ref{sec:seismicdirect} that including the 
%metallicities does not have a 
%significant effect on the derived \logg.
%in order to derive precise and accurate radii, masses, and ages, however, \feh\ 
%needs to be included --- this was also highlighed by \citet{gai11} with simulations.

%\red{BILL: Details on the selection of targets, duration of observations, and 
%extraction of seismic and photometric quantities (published anywhere???)}

\subsection{Seismic \logg\ from different sets of observations\label{sec:diffobs}}

\begin{figure}
\includegraphics[width=0.48\textwidth]{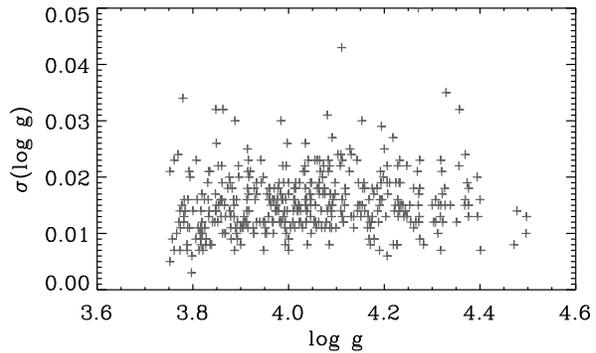}
\caption{Distribution of uncertainties in \logg\ $\sigma(\log g)$ as a function
of \logg\ for a sample of 403 \kep\ stars, using RadEx10 with 
observational constraints comprising \{$T_{\rm effPin}$,\mlsep,\numax\}.\label{fig:erlogg}}
\end{figure}

Using our validated method described in Sect.~\ref{sec:radex10}, we calculated 
values of \logg\ and their uncertainties for the sample of 403 stars
using the set of observations comprising \{\teff$_{\rm Pin}$,\mlsep,\numax\} 
while
adopting a mean \feh\ = -- 0.20 \pmm\ 0.30 dex.
We refer to this set as the reference set.   
The distribution of the uncertainties as a function
of \logg\ is shown in Figure~\ref{fig:erlogg}.
Here it can be seen that typical uncertainties in \logg\ for this set of 403
\kep\ stars is below 0.02 dex (there is one star with an error of 0.05 dex),
with a mean value of 0.015 dex.

\begin{figure}
\includegraphics[width=0.48\textwidth]{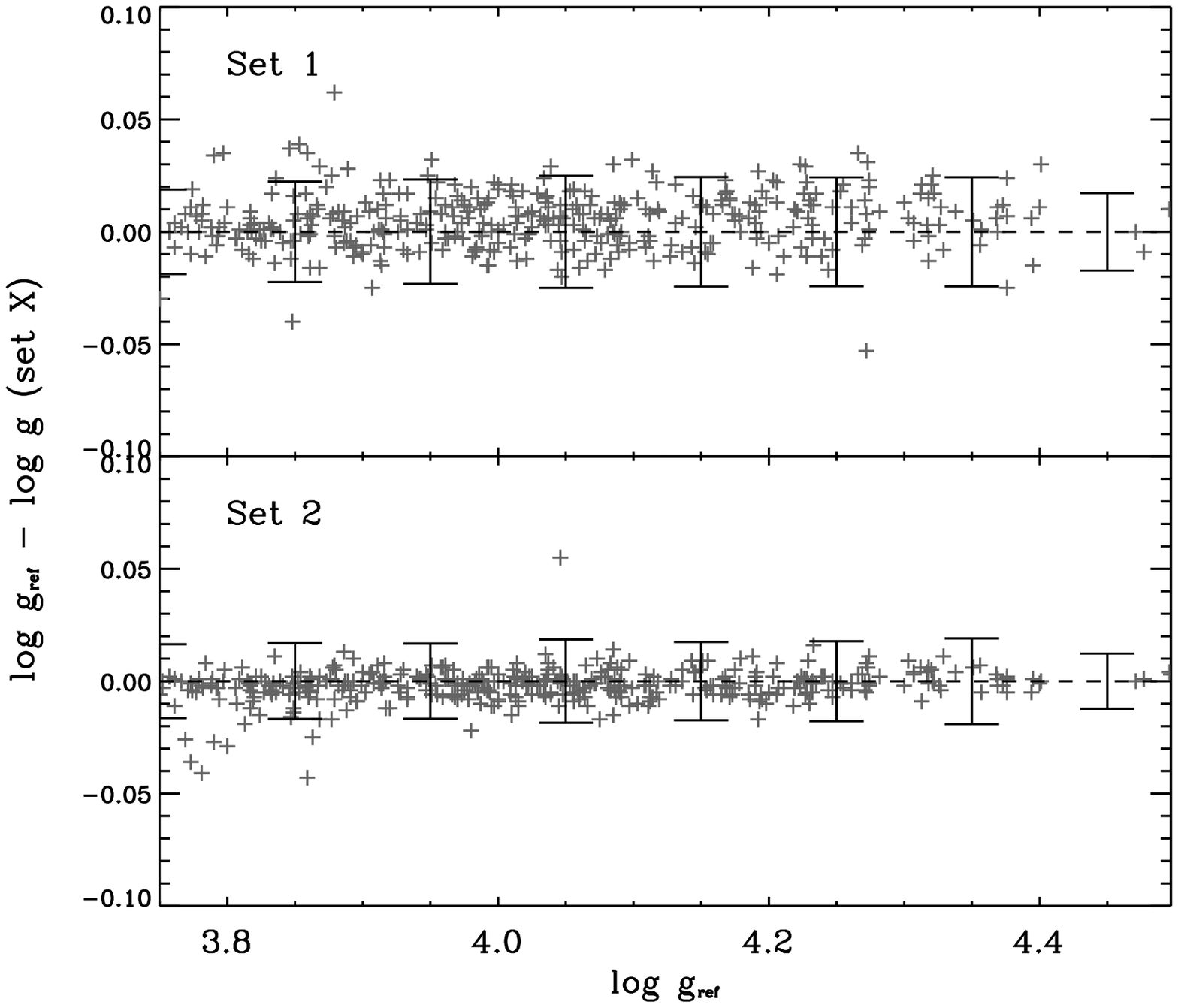}
\caption{Differences in \logg\ using subsets of observational constraints.
The reference set comprise (\mlsep, \numax, \teff$_{\rm Pin}$).
Set 1 and 2 comprise (\mlsep, \teff$_{\rm Pin}$) and 
(\mlsep, \numax, \teff$_{\rm IR}$), respectively.
For all sets \feh\ = --0.20 dex.
The $\pm 1\sigma$ error bars are average error bars measured over bins of 0.1 dex corresponding to Set X.  See Sect.~\ref{sec:diffobs} for details.
\label{fig:agreeset}}
\end{figure}

In Figure~\ref{fig:agreeset} we show the difference in the fitted \logg\ 
while considering different input observational sets compared to 
the reference set `log g$_{\rm ref}$'. 
%comprises the three original constraints; \mlsep, \numax, and
%\teff.  
The subsets are: Set 1 which considers \mlsep\ and \teff$_{\rm Pin}$ only and
Set 2 which %considers \numax\ and \teff\  only, 
%while Set 3 
considers \mlsep, \numax, and \teff$_{\rm irfm}$. 
%where \teff$_{\rm IR}$ is derived using 
%relations from the infra-red flux method \citep{cas10} 
%and implemented in \citet{sil12}. %\red{(Casagrande et al. in prep.)}.
We note that for all of the analyses 
\feh\ was constrained to  --0.20 \pmm\ 0.30 dex, which corresponds to $>$90\%
of the models of the grid.

Inspecting the top panel of Fig.~\ref{fig:agreeset} (Set 1)
one can see that by omitting \numax\ as an observed quantity can
result in 
differences of over {0.05 dex} for a very small percentage of the 
stars, but the absolute difference between the full set of 
results is +0.005 dex with an rms of 0.01 dex. 
Here, we note that several authors have investigated the relation between
\mlsep\ and \numax\ and find tight correlations \citep{bed03,stello09b}.
{The uncertainties arising from a set of data with less constraints
usually increases and we indeed find an increase in the uncertainties $\sigma$
of $\sim0.01$ dex.
The extra scatter of 0.01 dex is taken care of in the larger $\sigma$. }
%allows for the extra scatter. % is evident. 

Inspecting the lower panel of Fig.~\ref{fig:agreeset} (Set 2) 
one can see that the \teff\ 
derived by using different photometric scales 
results in a mean difference in \logg\ 
of -0.002 dex (i.e. no significant overall effect) with an rms scatter of 
0.007 dex.  This latter fact implies that we can
expect to add just under 0.01 dex to the error budget in \logg\ by considering 
\teff\ derived from different methods.  We found a similar result in 
Sect.~\ref{sec:systerrors} for $\beta$ Hydri.

%0.005, rms 0.013 set 2
%-0.005, rms 0.015 set 3 

%Inspecting the top two panels of Fig.~\ref{fig:agreeset} one can easily
%see that omitting either of the seismic quantities can cause a 
The mean value of the derived uncertainties ($\sigma$) in \logg\ for Set 1 and 2 are %, and 3 is 
%%below 0.03, 0.04, and 0.02 dex, respectively, with and the mean uncertainties
%%are 
0.023 and 0.015 dex, respectively, while those for the reference
set are 0.015 dex.  % for the 
%%reference set):
%omitting either of seismic indices increases the uncertainty by
%0.01 dex.   
The accuracy of these \logg\ (if we consider the reference set to be 
correct) is within a precision of 1$\sigma$.

\begin{figure}
\includegraphics[width=0.48\textwidth]{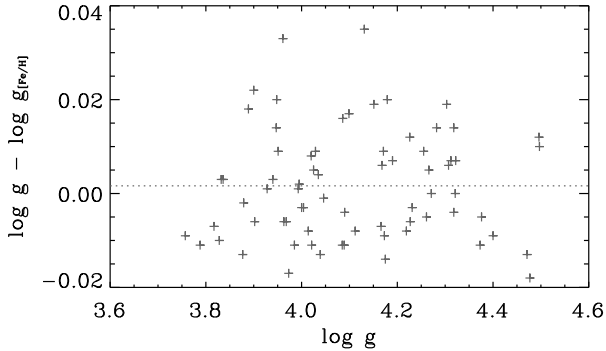}
\caption{Differences in the derived seismic \logg\ from RadEx10 
between adopting \feh\ = --0.20 \pmm\ 0.30
for all of the stars (\logg) and adopting [Fe/H] from 
spectroscopic analyses by \citet{bru12}, \logg$_{\rm [Fe/H]}$. 
See Sect.~\ref{sec:diffobs} for details. \label{fig:bruntt}}
\end{figure}

Figure \ref{fig:bruntt} compares the derived \logg\ using the 
reference set of observations, to those
with measured \teff\ and \feh\ from \citet{bru12}.  
The absolute mean residual is 
0.002 dex and is highlighted by the dotted grey line.  
We find that \logg\ can differ by up to 0.02 dex by including \feh.
This 0.02 dex is also consistent with what we found in 
Sect.~\ref{sec:systerrors} for $\beta$ Hydri when we considered different
metallicity constraints.  

%\red{numbers for these??}.

% for MS stars.
%For $\log g < 3.5$ we see that both Set 2 and Set 3 give similar 
%results (mean differences of 0.00 dex), 
%while for Set 1 a systematic shift in \logg\ that
%increases with decreasing \logg\ is very notable (up to 0.10 dex).  
%This same trend is found among all of the pipeline results.
%We also note that in general the error bars are larger at this end.
% and should
%account for most of the differences.

%\red{One would expect that using more input observational data would result
%in not only more precise values but also more accurate ones.  
%Any systematic error in one of the measurements would hopefully be compensated
%for by the other measurements.  
%For this reason we consider the results using the three constraints as
%the most accurate, and based on this fact, we can then conclude that 
%for most of the stars with 
%results based on only \mlsep\ and \teff\ or \numax\ and \teff\ are just
%as reliable but with uncertainties that increase somewhat.}
%%may not be as reliable and may suffer from a systematic shift of up to 0.10 dex.

%%%%%%%%%%%%%%%%%%%%%%%%%%%%%%%%%%%%%%%%%%%%% begin compare codes

\subsection{Comparison of results using different codes\label{sec:yrec}}
To investigate the possible source of systematics arising from 
using a different evolution code and input
physics, we determined \logg\ using a second pipeline code, Yale-Bham
\citep{gai11,cre12kep}.  
Details of the code can be found in the cited papers.  Here, it suffices
to know that the method is very similar to that of RadEx10, but the 
evolution code is based on YREC \citep{dem08} in its non-rotating configuration,
with the following specifications:  OPAL EOS \citep{rn02} and OPAL high-temperature opacities \citep{ir96} supplemented with low-temperature 
opacitites from \citet{fer05}, and the NACRE nuclear reaction rates \citep{ang99}.  Diffusion of helium and heavy-elements were included, unlike RadEx10.
%\red{details of opacities etc.}.

\begin{figure}
\includegraphics[width=0.48\textwidth]{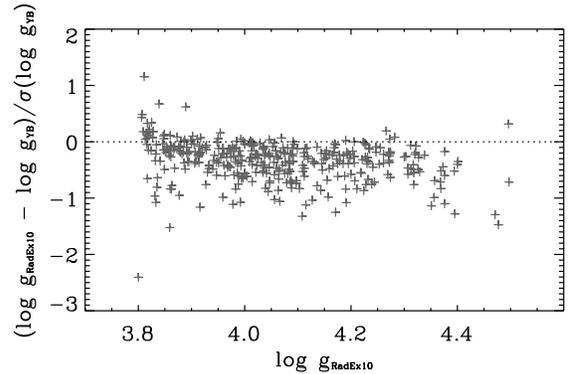}
\caption{The difference between the values of \logg\ obtained by RadEx10 
and Yale-Bham as a function of Radex10 \logg\ 
and normalised by the Yale-Bham errors.
See Sect.~\ref{sec:yrec} for details.
%  The error bars are the 
%mean values obtained for each bin of 0.10 dex in \logg.
\label{fig:agreemod1} }
\end{figure}

Figure~\ref{fig:agreemod1} shows the difference in 
the derived value of \logg\ between RadEx10 and Yale-Bham using 
\{\mlsep,\numax,\teff$_{\rm Pin}$,\feh=--0.2\}. 
We show the normalised difference, i.e. divided by the Yale-Bham errors, which
are very similar to the RadEx10 errors.
%The error bars that are overplotted are the mean values of the errors for 
%that range of \logg\ for YREC (similar to RadEx10 error bars).
As can be seen from the figure the agreement
in \logg\ between the different methods is within  
1$\sigma$ or $<0.01$ dex for 98\% 
of the stars and 2$\sigma$ for all stars except one.
%($<0.02$ dex for all stars).
A small absolute difference of 0.005 is found with 
an rms of 0.005 (units of $\sigma$).
This is most likely due to the different physics adopted in the codes.
A very slight systematic trend is present with a slope of -0.005 \pmm\ 0.005.
%The systematic is most likely due to
%%but this systematic is on the order of or less than 0.01 dex, 
%%due most likely to 
%the different sets of physics and the evolution models used.
The agreement between the results using the different codes and methods
is very encouraging.
We can consider an upper limit of 0.01 dex as a typical systematic 
error arising from employing different codes.
%\red{Is it still necessary to discuss differences using sarbani's other
%results? if so, I can include a few sentences.}

%We used three pipeline codes to analyse the same set of data. 
%These codes are RadEx10 (Sect.~\ref{sec:radex10}),  RADIUS \red{cite}, and
%\red{Sarbani's, citation}.  
%These three codes have also been presented and compared
%in \citep{cre12kep}, and in Table~\ref{tab:codedifferences} 
%we summarize the differences between
%them. \red{include table here}.
%%Details of some of the codes were presented in \citet{cre12} and 
%%a summary of the input physics, evolution codes, and methods is given
%%in the Appendix \red{perhaps include the table from the wiki?}.

%Each of the pipeline codes
%are similar to that presented in Sect.~\ref{sec:radex10}.
%Variations in the physics include different equations of state, tables
%of opacities, treatment of the atmosphere and mixing-length theory.
%Several stellar evolution codes were used e.g. ASTEC, YREC. 
%The methods for choosing the optimal stellar property also varied --- 
%some studied a distribution of perturbed observations, others chose an
%optimal model from the grid (see Table~\ref{tab:codedifferences}).
%All of the methods used \teff, \mlsep, and \numax\ as the input 
%observations and derived a value of \logg\ with an uncertainty.

%\begin{table*}
%\begin{center}\caption{Summary of pipeline codes used to determine \logg.\label{tab:codedifferences}}
%\begin{tabular}{lllllll}
%\hline\hline
%Pipeline & Stellar Evolution Code & Equation of state & \\ 
%\hline
%To be filled in\\
%\hline\hline
%\end{tabular}
%\end{center}
%\end{table*}

%\subsection{\logg\ determined by different methods}
%All of the modellers used \mlsep, \numax, and \teff\ as the input constraints
%to their codes in order to determine \logg. 
%Most of the grid-based methods explore a larger range of evolutionary 
%stages, i.e. as far as \logg\ = 3.0, while
%RadEx10 was initially developed for studying only main sequence stars. 
%For the latter, we consider only the results for $\log g > 3.8$ (XX stars) ---
%this range is also valid for the reference stars studied in Sect.~\ref{}.
%The value of \logg\ determined by all of the methods are in 
%very close agreement, and

%In Figure~\ref{fig:agreemod1} we show the differences in the derived 
%\logg\ between
%RadEx10 and RADIUS and RadEx10 and YY for the sample of 378 stars with \logg\ 
%$>$ 3.8 (range considered in Sect.~\ref{sec:section3}).

%%%%%%%%%%%%%%%%%%%%%%%%%%%%%%%%%%%%%%%%%%%%% end compare codes

\begin{figure}
\includegraphics[width=0.48\textwidth]{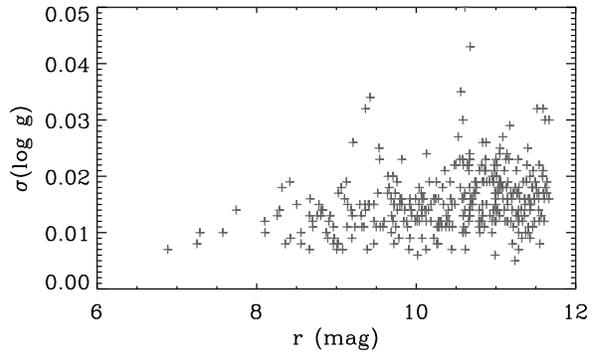}
\caption{Distribution of derived uncertainties in the seismic \logg\ 
from RadEx10 for 403 
\kep\ V/IV stars 
as a function of SDDS r magnitude.
\label{fig:ermag}}
\end{figure}

\subsection{Uncertainties and systematic errors 
in the seismic \logg\ as a function of magnitude}

%Considering the reference set of input data (\mlsep, \numax, and \teff), 
%in Figures
%Assuming that both seismic quantities and \teff\ are available, in Figures 
In Figure \ref{fig:ermag} %; and \ref{fig:erlogg}
we show the uncertainties in \logg\ as a function of SDDS r magnitude
(obtained from the KIC) while using the reference set of observations.  
As the figure shows, precision improves with apparent brightness, where
more reliable measurements have smaller error bars.
%
%As expected, Fig.~\ref{fig:ermag} shows that 
%precision improves with apparent brightness, where the 
%measurements are much more reliable, and hence have smaller errors.
We can expect typical uncertainties in \logg\ of less
than 0.03 dex for the full sample with a  
mean value of $<0.02$ dex.
%In this figure the subset of stars
%with $\log g < 3.5$ are shown by the black boxes and do not show
%any difference in trend than the MS stars.

{In Table~\ref{tab:kepunc} we summarize for different apparent 
magnitude bins 1) the mean uncertainties $\langle \sigma \rangle$ and 
2) the mean systematic errors $\langle s \rangle$ arising from adopting different temperature
scales $\langle s_{T{\rm eff}}\rangle$, metallicities $\langle s_{\rm{[Fe/H]}} \rangle$ and different
pipeline codes $\langle s_{\rm code} \rangle$.
Note that the sample size for the bin with $r<9$ is much smaller than 
the others and this is the reason we find a larger $s_{\rm [Fe/H]}$ for 
$r<9$.
In the last row we give the mean uncertainty and the 
adopted systematic errors over the full sample.}

\subsection{Summary of results for the sample of \kep\ stars}
From the sample of \kep\ data with  $3.75 < \log g < 4.50$ derived
from seismic data, we obtain  
%presented here, 
%we conclude that the sample 
a mean precision $\sigma$ of between 
0.01 and 0.02 dex (max 0.03 dex for 99\% of stars) 
%in \logg\ for $3.75 < \log g < 4.50$ 
when \mlsep, \numax, and \teff\ are used
as the input constraints for the pipeline analysis. 
A typical systematic error $s$ of 0.01 dex can be added to account for
a possible systematic error in \teff,
such as that obtained by applying different calibration methods to photometry.
Similarly, we can add a systematic error of 0.02 dex due to 
an incorrect or lack of a \feh\ measurement. 
We also showed that using 
different models and physics in the pipeline analyses yields results 
in \logg\ consistent to within 0.01 dex i.e. almost no model dependence.
%This gives an upper limit of 0.04 dex to account for systematic errors.
%In Table~\ref{tab:kepunc} we summarize this information while taking 
%into account the observed magnitude of the star.

%If only one of the seismic datum are available, we should be cautious
%using mlsep\ and \teff\ alone for stars with \logg $< 3.4$ dex, as 
%differences of up to 0.10 dex can be found.
Eliminating one of the seismic indices from 
the set of data yields a zero offset 
%%of (\numax, \teff) or (\mlsep,\teff) 
%%compared to (\mlsep, \numax, \teff) 
%does not reveal any systematic differences 
in the results for the full sample
of stars and a typical scatter of 0.01 dex,  although differences of over 0.05 dex
were found for $<1\%$ of the stars.
%, and a typical scatter of 0.01 dex
%was evident.  %However, the uncertainties also increased by 0.01 dex.

%n general the error bars are also larger by about 0.01 dex for all stars
%when using only one of either \mlsep\ and \numax\ along with \teff.
%%can beexpected due to the different \teff\ measurements.  
%%This is similar to what we concluded in Sect. for $\beta$ Hydri.

%\pagebreak

\begin{table}
\begin{center}
\caption{Summary of uncertainties $\sigma$ 
and systematic errors $s$ in the seismic \logg\ 
from the \kep\ sample of stars as a function of magnitude.
We give the mean values $\langle \rangle$ of each over the sample that falls into
the corresponding magnitude bin.
The systematic errors $s$ have subscripts which refer to possible sources
of errors in the input \teff, \feh, and the code that is used to generate
the stellar models (see Sects.~\ref{sec:diffobs} and \ref{sec:yrec} for details).  In the final row we summarize the mean uncertainties and adopted systematic
errors for the full sample.
\label{tab:kepunc}}
\begin{tabular}{lcccccccccccc}
\hline\hline
r & $\langle \sigma \rangle$ & $\langle s_{\rm Teff} \rangle$ & 
$\langle s_{\rm [Fe/H]} \rangle$ & $\langle s_{\rm code} \rangle$ \\
(mag) & (dex) & (dex) & (dex) & (dex)\\
\hline
%$<9$ & \\
%$9<V<10$ &\\
%$10<V<11$ &\\
%$11<V<12$ &\\
 $6<{\rm r}<$9&0.012&0.004&0.014&0.005\\
$9<{\rm r}<10$&0.014&0.004&0.009&0.004\\
$10<{\rm r}<11$&0.016&0.009& ...&0.006\\
$11<{\rm r}<12$&0.017&0.009&...&0.006\\
\\
$6<{\rm r}<12$ &0.015 & 0.01 & 0.02 & 0.01\\
\hline\hline
\end{tabular}
\end{center}
\end{table}

\section{Discussion}

The first objective of this study was to investigate if we can determine
\logg\ reliably using global seismic quantities and atmospheric data.  
We showed that our method is reliable by comparing our results with 
values of \logg\ derived from direct mass and radius estimates of seven bright
nearby stars.
We then applied our method to a list of 40 Sun-like stars and derived
\logg\ to within 1$\sigma$ ($\sim0.01$ dex) 
of those from \citet{mat12} and \citet{sil12}.  
We also showed that for a sample of 400+ \kep\ stars (6 < r mag < 12)
with \logg\ between 3.75 and 
4.50 dex typical uncertainties $\sigma$ of less than 0.02 dex can be expected and
we estimated systematic errors $s$ of no more than 0.04 dex arising
from errors in \teff\ and \feh\ measurements and using different codes.

All of the \kep\ stars will be observed by the Gaia mission, and for this
reason they provide a valid set of calibration stars, by 
constraining \logg\ with precisions and accuracies 
much better than spectroscopic or isochrone methods provide for the current
list of calibration stars for Gaia.
The astrophysical parameters to be determined from Gaia data using the 
astrometry, photometry and BP/RP spectrophotometry are \teff, \ag, 
\feh, and \logg.  By ensuring that the \gspp\ methods deliver \logg\ consistent
with the seismic value will reduce the uncertainties and inaccurcies in 
the other parameters.
Moreover, an independent \logg\ provides an extra constraint 
for the determination of the atmospheric parameters from \gsps\ where
degeneracies between \teff\ and \logg\ severely inhibit the precision 
of atmospherically extracted parameters and chemical abundances.

%\red{kepler field for distance (da Silva)
While in this paper we concentrated primarily on using \kep\ data to 
determine \logg, we note that
both the \corot\ and \kep\ fields have great potential in other aspects. 
For example, \citet{mig12epjwc} and \citet{mig12mnras2} determine distances, 
masses, and ages of red giants from
the CoRoT fields to constrain galactic evolution models, while
\citet{sil12} develop a method which
couples asteroseismic analyses with the infrared flux method to 
determine stellar parameters including effective temperatures and 
bolometric fluxes (giving angular diameters), and hence distances.  
%, although we note that 
%\citet{sand12} suggest systematic errors on masses of giants from
%asteroseismic predictions are possible.
The distances obtained by both  \citet{mig12mnras2} and \citet{sil12}
can be compared
directly with a Gaia distance for either calibration of Gaia data or stellar 
models.
%In particular, \citet{sand12} suggest systematic errors on masses of giants 
%from asteroseismic predictions are possible.
By combining data from \corot\ and \kep\ with 
those from Gaia we can also determine extremely precise 
angular diameters by coupling a seismically determined 
radius with the Gaia parallax.
Finally, these stars will also be excellent calibrators for 
the FLAME workpackage of Gaia, which aims to determine masses and ages
for one billion stars.

%-Fixing \logg\ in \gspp.... typical errors.... impact for GSP\_spec too.
%-precise logg for abundance determinations
%\red{-using a Gaia distance + seismic logg.}

\section{Conclusion}

In this work we explored the use of %of the mean seismic quantities observed
F, G, K IV/V stars observed in the \kep\ field as a possible source
of calibration stars for fundamental stellar parameters from the Gaia mission.
Our first objective was to test the reliability/accuracy of a seismically
determined \logg, and using a sample of seven bright nearby targets we
proved that seismic methods are accurate (see Fig.~\ref{fig:logg})
by obtaining results to within 0.01 dex of the currently accepted
\logg\ values.
%We also determined a seismic radius and mass for these stars and 
%these are presented in Table~\ref{tab:radexgmr}.
%These also yield better than 1$\sigma$ consistency with accepted values.
We showed, however, that the accuracy of the input atmospheric parameters
does play a role in the accuracy of the derived parameters.  For 
$\beta$ Hydri we found that errors in the atmospheric parameters, and in
particular \feh, can 
change \logg\ by 0.02 dex. % and by much more for radius and mass determinations.

%In particular, we showed that the seismic data deliver not only 
%precise determinations of \logg\ \citep{mm11,gai11,cre12kep} but using
%a sample of bright nearby targets, we showed that grid-based methods
%also provide {\it accurate} determinations of \logg\ (see Fig.~\ref{fig:logg}).
%We determined a seismic \logg, radius, mass of a sample of bright nearby
%targets and these are presented in Table~\ref{tab:radexgmr}.  
%Our results are consistent to less than 1$\sigma$ of the currently accepted
%values.
%These results support the use of a {\it seismic \logg}.
%We showed, however, that the accuracy of the input atmospheric parameters
%does play a role in the accuracy of the derived \logg.  For 
%$\beta$ Hydri we showed that errors in the atmospheric parameters can 
%change \logg\ by 0.02 dex.

We then applied our grid-based method RadEx10 to an extended sample
of stars from the literature.
%which have been studied both seismically and atmospherically in the literature.  
We derived their seismic \logg\ 
%stellar properties (\logg, $R$, $M$, $L$, age) and 
and these are given in Table~\ref{tab:derivedextlist}.  
We showed that a grid-based
\logg\ is consistent with the values of \logg\ obtained by detailed
seismic analysis from \citet{mat12} for 22 stars 
and in excellent agreement with 
the results from \citet{sil12} for the 6 common stars.
We find typical precisions in \logg\ of  0.02 dex.

We finally studied the distribution of errors in \logg\ from  
two analysis methods for a sample of 403 \kep\ stars with $3.75<\log g<4.5$, 
and we obtained a typical uncertainty of between 
0.01 and 0.02 dex in \logg\ 
for F, G, K IV/V stars (see Table~\ref{tab:kepunc}).
We can add a total of 0.04 dex as a systematic
error which arises from the adopted temperature scales (0.01 dex) and measured
metallicities (0.02 dex) as well as the 
grids of models used (0.01 dex), which differ in evolution codes and input
physics. %(different codes and physics). 
The precisions in the data are unprecedented, and the systematic errors
are much smaller than those stemming from any other method, 
especially for $7<V<12$ stars
(see e.g. \citealt{cre12kep} Fig~1 and 
Table~3 which compare spectroscopically derived
\logg\ for five \kep\ stars with $V\sim11$).

As a final remark, we highlight that while there are $>$500 IV/V stars in 
the   \kep\ field and some in the CoRoT fields that exhibit Sun-like oscillations, 
there are also 
1000's of red giants in both CoRoT and \kep\ fields with these same measured
quantities, however, the 
accuracy of these
%Exploring the accuracy of these
stars' seismic \logg\ is yet to be studied.

\section*{Acknowledgements}
OLC thanks Luca Casagrande and Victor Silva Aguirre for making
data available.
SB acknowledges NSF grant AST-1105930.
%AMS is  partially supported by the Marie Curie International Reintegration Grant  PIRG-GA-2009-247732 within the 7$^{\rm th}$ European Community Framework Programme, and by the  MICINN grant  AYA2011-24704.
AMS is  partially supported by the European  Union International Reintegration
Grant  PIRG-GA-2009-247732,  the  MICINN   grant  AYA2011-24704,  by  the  ESF
EUROCORES Program  EuroGENESIS (MICINN grant EUI2009-04170), by  SGR grants of
the Generalitat de Catalunya and by the EU-FEDER funds.
WJC and YE thank the UK STFC for grant funding to support asteroseismic research.
MJPFGM acknowledges the support through research grant {\small PTDC/CTE-AST/098754/2008}, from FCT/MEC~(Portugal) and FEDER~(EC). 
OLC is a Henri Poincar\'e Fellow at the Observatoire de la C\^ote d'Azur (OCA),
funded by OCA and the Conseil G\'en\'eral des Alpes-Maritimes.

\bibliographystyle{mn2e}
\bibliography{ref}

\end{document}